\documentclass[journal]{IEEEtran}

\usepackage{cite}
\usepackage{graphicx}
\usepackage{amssymb,amsmath,bm}
\usepackage[colorlinks,linkcolor=black,anchorcolor=black,citecolor=black,urlcolor=black]{hyperref}
\usepackage{hyperref}
\usepackage{multirow}
\usepackage{amsfonts}
\usepackage{url}
\usepackage{multirow}
\usepackage{color}
\usepackage{tabularx}
\usepackage{color}
\usepackage{algorithm}
\usepackage{algpseudocode}
\usepackage{diagbox}
\usepackage{comment}
\usepackage{enumitem}
\usepackage{booktabs}

\ifCLASSINFOpdf
\else
\fi
\begin{document}

\title{CFMDCTCodec: A Low-Bitrate Neural Speech Codec with Noise-Prior-aware Conditional Flow Matching for MDCT-Spectral Enhancement}

\author{Xiao-Hang~Jiang,~Yang~Ai,~\IEEEmembership{Member,~IEEE},~Hui-Peng~Du,~Zhen-Hua~Ling,~\IEEEmembership{Senior Member,~IEEE},~Ji~Wu,~\IEEEmembership{Senior Member,~IEEE}
\thanks{This work was funded by the National Nature Science Foundation of China under Grant 62301521. (Corresponding author: Yang~Ai)}
\thanks{Xiao-Hang Jiang, Yang Ai, Hui-Peng Du and Zhen-Hua Ling are with the National Engineering Research Center of Speech and Language Information Processing, University of Science and Technology of China, Hefei, 230027, China (e-mail: jiang\_xiaohang@mail.ustc.edu.cn, yangai@ustc.edu.cn, redmist@mail.ustc.edu.cn, zhling@ustc.edu.cn).}
\thanks{Ji Wu is with the Department of Electronic Engineering, Tsinghua University, Beijing, 100084, China (e-mail: wuji\_ee@tsinghua.edu.cn).}
}
\markboth{}%
{Shell \MakeLowercase{\textit{et al.}}: Bare Demo of IEEEtran.cls for Journals}
\maketitle

\addtolength{\textfloatsep}{-0.2cm}
\addtolength{\dbltextfloatsep}{-0.2cm}
\begin{abstract}
High-quality speech coding at low bitrates is crucial for bandwidth-constrained applications, yet remains challenging due to the severe loss of quality-critical information in highly compressed representations. 
To overcome this challenge, we propose CFMDCTCodec, a low-bitrate neural speech codec that operates entirely in the modified discrete cosine transform (MDCT) domain.
CFMDCTCodec integrates a lightweight encoder–quantizer–decoder–style MDCT-spectral codec with a noise-prior-aware, conditional-flow-matching (CFM)-based MDCT-spectral enhancer.
Within this framework, the codec serves as a base module that compactly discretizes the MDCT spectrum extracted from speech and produces an initial coarse reconstruction, while the enhancer further restores fine-grained spectral details.
The enhancer improves the decoded MDCT spectrum by integrating a conditional MDCT velocity-field filter with an ordinary differential equation (ODE) solver, under the guidance of an MDCT-derived magnitude-adaptive noise prior, aiming to emphasize perceptually significant high-energy regions while stabilizing low-energy and silent regions. 
Finally, the enhanced MDCT spectrum is reconstructed into the decoded speech using the inverse MDCT.
When optimizing CFMDCTCodec, we adopt a unified non-adversarial training strategy that jointly combines reconstruction, quantization and CFM objectives. 
Both objective and subjective evaluations show that CFMDCTCodec outperforms competitive baselines in low-bitrate regimes, e.g., 0.65~kbps, while approaching the perceptual quality of large-scale codecs with significantly fewer parameters and computations.
\end{abstract}
\IEEEpeerreviewmaketitle

\section{Introduction}

\IEEEPARstart{S}{peech} codecs map a speech waveform to a compact bitstream and reconstruct it at the decoding stage, trading bitrate against perceptual quality and computational cost \cite{noll1997iso, tremain1976linear, kroon2003regular}.
Conventional telecommunication and streaming systems typically operate at several kilobits per second (kbps) \cite{o2002linear, salami2002toll, valin2013high, dietz2015overview}, which suffices for mobile networks and internet voice services. 
By contrast, emerging applications such as satellite and high-frequency radio links and large-scale cloud-based speech monitoring operate under extremely tight bandwidth and energy budgets, where even a few hundred bits per second (bps) per stream is costly.
These scenarios motivate speech coding in the low-bitrate regime, where only a handful of latent symbols can be transmitted per second. 
Consequently, speech decoding becomes highly ill-posed, as the severely limited bitstream preserves only coarse structure while fine-grained details that shape naturalness and speaker traits cannot be explicitly conveyed. 
Therefore, achieving high-quality speech coding at low bitrates poses a significant challenge, yet overcoming it is of substantial practical importance.

Early traditional speech codecs such as adaptive multi-rate and adaptive multi-rate wideband (AMR/AMR-WB) \cite{bessette2003adaptive}, enhanced voice services (EVS) \cite{dietz2015overview}, and Opus \cite{valin2012definition} primarily relied on carefully engineered digital signal processing (DSP) pipelines that combine linear prediction, code-excited linear prediction (CELP) \cite{schroeder1985code}, and psychoacoustic modeling.
They are highly optimized within their design ranges but degrade rapidly when pushed to low bitrates, often exhibiting strong artifacts and loss of naturalness. 

With the advent of deep learning, neural speech codecs {\cite{zeghidour2021soundstream, defossez2023high, wu2023audiodec, kumar2024high, jiang2024mdctcodec, xin2024bigcodec, welker2025flowdec, yang2024generative, san2023discrete, pia2025flowmac, liu2024semanticodec, xu2025mucodec}} have enabled a more favorable tradeoff between perceptual quality and bitrate by learning an end-to-end neural encoder–quantizer–decoder architecture directly from training data. 
Broadly, existing approaches can be grouped by their modeling domain into waveform-based and spectral-based neural speech codecs. 
Waveform-based codecs such as SoundStream \cite{zeghidour2021soundstream} and follow-up works including EnCodec \cite{defossez2023high}, AudioDec \cite{wu2023audiodec}, and DAC \cite{kumar2024high} can deliver high-quality speech at a few kbps, typically leveraging adversarial training to promote perceptual realism.
These codecs typically adopt residual vector quantization (RVQ) \cite{juang1982multiple} for latent discretization. 
In RVQ, multiple codebooks are applied sequentially, with each stage quantizing the residual left by previous stages, thereby progressively refining the discrete representation. 
Although multi-stage quantization enhances the expressiveness of the bitstream and helps preserve fine-grained details, it also makes further bitrate reduction difficult in RVQ-based codecs, as lowering the bitrate requires shrinking the discrete capacity and often leads to noticeable quality degradation.

Beyond waveform-based methods, spectral-based neural speech codecs have recently attracted increasing attention as an alternative design point.
Compared with waveform-based codecs that directly model raw samples and demand higher computation and heavier training recipes, spectral-based codecs leverage time–frequency representations to exploit spectral structure, enabling lighter encoder–decoder backbones and more stable training. 
For example, in our prior work, we proposed MDCTCodec \cite{jiang2024mdctcodec}, which operates directly on real-valued modified discrete cosine transform (MDCT) coefficients and adopts a compact, fully convolutional architecture. 
By working in the MDCT domain, MDCTCodec significantly reduces model size and computational complexity compared with waveform-based codecs.
However, MDCTCodec likewise relies on a residual quantization strategy and therefore suffers from the same discrete-capacity bottleneck at low bitrates.

A straightforward approach to achieving high-quality speech coding at low bitrates is to adopt a simple quantization scheme while enhancing the decoding capability. 
Along this line, one possible approach is to substantially increase the capacity of both the encoder and decoder, enabling the model to better learn the mapping from an aggressively quantized low-rate representation to a high-quality waveform.
BigCodec \cite{xin2024bigcodec}, for example, employs a single vector quantizer (VQ) scheme and significantly enlarges the encoder and decoder to {over a hundred million} parameters, incurring high floating-point operations per second (FLOPs) and model size while achieving strong reconstruction quality.
However, the resulting computational and model-size overhead limits practical deployment and runs counter to the lightweight design philosophy underlying spectral-domain codecs such as MDCTCodec.

Relying excessively on large models to achieve low-bitrate coding is not cost-effective. 
An alternative approach is to keep the original encoder–quantizer–decoder architecture unchanged and introduce a post-processor that further enhances the decoded output, which can be implemented using a lightweight generative model.
FlowDec \cite{welker2025flowdec}, for instance, employs a post-processor based on conditional flow matching (CFM) \cite{lipman2023flow} in the short-time Fourier transform (STFT) domain, which refines the decoded speech in the time–frequency space and then resynthesizes it back to the waveform domain.
{While this approach enables FlowDec to achieve high-perceptual-quality speech reconstruction, it primarily focuses on higher bitrate settings.}

To address the challenge of high-quality and low-bitrate speech coding, {inspired by \cite{welker2025flowdec, wu2024scoredec, jiang2024mdctcodec}}, we propose CFMDCTCodec, which combines a single-codebook MDCT-spectral codec with a noise-prior-aware CFM-based MDCT-spectral enhancer, building on our prior MDCTCodec \cite{jiang2024mdctcodec}. 
All components of CFMDCTCodec operate entirely in the MDCT spectral domain, fully exploiting the efficiency advantages of spectral-domain modeling. 
The single-codebook MDCT-spectral codec aggressively compresses the MDCT spectrum extracted from speech to achieve a low-bitrate representation and decodes it into a coarse MDCT spectrum, which is subsequently refined by a decoder-side enhancer. 
To ensure stable MDCT-domain enhancement based on CFM, we first apply range normalization to the coarse spectrum and then construct a magnitude-adaptive noise prior derived from its spectral energy. 
Starting from the noise prior and conditioned on the normalized coarse MDCT spectrum, CFM traces a trajectory that progressively yields a higher-quality MDCT spectrum. 
Finally, the enhanced MDCT spectrum is converted back into the decoded speech waveform via inverse MDCT (IMDCT). 
CFMDCTCodec is trained end-to-end via joint optimization of the MDCT-spectral codec and enhancer under combined reconstruction and CFM objectives. 
Both objective and subjective evaluations confirm that the proposed CFMDCTCodec achieves significantly better decoded speech quality than baseline codecs at low bitrates (e.g., 0.65 kbps), while requiring lower model complexity and computational cost. 
{Speech samples are available on our demo page\footnote{Speech examples are available at \href{https://xhjiang1.github.io/CFMDCTCodec}{https://xhjiang1.github.io/CFMDCTCodec}.}.}

The contributions of CFMDCTCodec are threefold. 
First, {it} introduces a new solution for low-bitrate speech coding in the MDCT spectral domain by combining a single-codebook compression with decoder-side post-processing enhancement. 
Second, tailored to the characteristics of MDCT representations, {it} develops a coarse-to-fine CFM-based MDCT enhancement strategy with MDCT normalization and a magnitude-adaptive noise prior, enabling effective compensation for distorted spectra. 
Third, {it adopts} an end-to-end joint training strategy for the single-codebook codec and the CFM-based enhancer, avoiding adversarial training and enabling simpler and more efficient learning.

The rest of this paper is organized as follows. 
Section~\ref{sec:related} reviews prior work on spectral-based neural speech codecs and flow-matching-based generative models used for speech processing. 
Section~\ref{sec:method} describes the details of the proposed CFMDCTCodec. Section~\ref{sec:experiment} presents the experimental setup and results. 
Finally, Section~\ref{sec:conclusion} concludes the paper and discusses potential directions for future work.

\section{Related Work}
\label{sec:related}
\subsection{Spectral-based Neural Speech Codecs}
\label{subsec:rw_spectral}

Recent work has increasingly explored speech codecs that operate on time–frequency representations, which is also the focus of this paper. 
Compared to waveform-based approaches that directly model raw samples and often require heavy computation and complex training recipes, spectral-based codecs leverage structured time–frequency representations and typically enable lighter architectures and more stable optimization. 
These spectrum-based codecs generally rely on invertible time–frequency transforms and can be broadly categorized into STFT-based and MDCT-based approaches. 

For STFT-based codecs, a key challenge lies in handling phase information. 
APCodec \cite{ai2024apcodec} addresses this issue by explicitly modeling both amplitude and phase in parallel branches, leveraging neural phase prediction technique to improve reconstruction fidelity. 
ComplexDec \cite{wu2025complexdec} further extends STFT-based coding to the complex spectrum by jointly modeling real and imaginary components, thereby avoiding explicit phase modeling, but requiring a relatively large and computationally demanding backbone. 
Overall, STFT-based codecs must explicitly model or compensate for phase information, which often leads to increased model size and computational overhead, motivating alternative spectral representations such as MDCT.

{The MDCT provides a real-valued, critically sampled, overlapping time–frequency representation with excellent energy compaction, making it particularly attractive for lightweight neural speech codecs. Building on these properties, our previous work MDCTCodec \cite{jiang2024mdctcodec} operates directly in the MDCT domain, employing fully convolutional networks and RVQ. While achieving competitive quality at moderate bitrates, its performance degrades under severe compression. }

{To better clarify the connection between CFMDCTCodec and MDCTCodec \cite{jiang2024mdctcodec}, CFMDCTCodec preserves the same MDCT-domain representation and lightweight fully convolutional backbone as MDCTCodec, but incorporates several key modifications to better suit low-bitrate coding. 
Specifically, it adopts a single-codebook quantizer with forced-updating training, a CFM-based enhancer to restore severely compressed details, and a fully non-adversarial end-to-end joint training scheme. 
Detailed architectural descriptions are provided in Section \ref{sec:method}.}

\subsection{{Diffusion Models for Speech Codecs}}
\label{subsec:rw_diffusion}
{Diffusion models \cite{ho2020denoising, song2021score} have recently become powerful generative tools in speech processing, owing to their ability to refine degraded signals without adversarial training.}

{To address the perceptual artifacts common in low-bitrate neural speech codecs, recent studies have integrated diffusion models into the codec architecture. 
For instance, ScoreDec \cite{wu2024scoredec} uses a score-based diffusion post filter in the complex spectral domain to effectively restore missing high-frequency details and phase information, completely bypassing unstable adversarial training. 
Similarly, LaDiffCodec \cite{yang2024generative} utilizes a latent diffusion model as a generative de-quantizer to reconstruct high-fidelity continuous latent vectors from quantized tokens. Alternatively, Multi-Band Diffusion \cite{san2023discrete} directly replaces the generative adversarial network (GAN) based decoder with a diffusion vocoder, partitioning the audio into independent frequency bands to mitigate metallic artifacts without error accumulation.}

{However, standard diffusion models typically require computationally expensive iterative sampling, thereby motivating increasing interest in flow-matching-based alternatives that offer simpler training and faster inference for speech codecs.}

\subsection{Flow Matching for Speech Codecs}
\label{subsec:rw_flowmatching}

Flow matching has recently emerged as an alternative formulation to diffusion-based generative models. 
Rather than learning a score function or a reverse-time stochastic differential equation (SDE) {as in score-based diffusion models \cite{song2021score},} 
flow matching directly learns a deterministic velocity field that transports a simple reference distribution to the target data distribution by integrating an ordinary differential equation (ODE).
Lipman \MakeLowercase{\textit{et al.}} \cite{lipman2023flow} introduced flow matching for generative modeling, showing that regressing a deterministic velocity field enables competitive image generation without explicit likelihood estimation or reverse-SDE simulation. 
Subsequent work proposed refinements such as conditional and joint flow matching \cite{tong2024improving} and rectified flows \cite{liu2023flow}, further improving training stability and sampling efficiency. 

Flow matching has first seen broad adoption in the vision domain, where it has been successfully applied to high-fidelity and high-resolution image generation and editing \cite{esser2024scaling, liu2023flow}. 
In these settings, the velocity field is learned over pixel spaces or latent feature spaces, and the backbone is typically a U-Net \cite{ronneberger2015u} or a transformer \cite{vaswani2017attention} operating on two-dimensional feature maps. 
The flow objective is often combined with architectural techniques such as multi-resolution attention and positional encodings, many of which were originally developed for diffusion models. 

Building on its success in the vision domain, flow matching has gradually been extended to audio and speech processing applications. 
In these domains, CFM has begun to serve as a core building block for generative models. 
For example, F5-TTS \cite{chen2025f5} employs a CFM module to reconstruct speech from text by learning a velocity field in a latent acoustic space conditioned on linguistic and prosodic features. 
Beyond text-to-speech, other studies \cite{luo2025wavefm, liurfwave, jung2024flowavse, lee2025flowse} have explored flow-matching-based vocoders or postfilters operating on spectral representations to enhance perceptual quality, while avoiding the training instability commonly associated with adversarial approaches.

Recent work has also begun to explore the use of flow matching in neural speech codecs. 
FlowDec \cite{welker2025flowdec} proposes a two-stage neural coding framework that combines a waveform-domain base codec with a CFM-based postfilter operating in the STFT domain. 
Specifically, FlowDec first trains a non-adversarial waveform codec, i.e., a DAC \cite{kumar2024high} without adversarial components and then freezes it, after which a CFM is trained to refine the STFT spectra of the decoded speech. 
{This innovative design demonstrates the strong potential of flow matching for high-perceptual-quality audio reconstruction without adversarial training. However, FlowDec \cite{welker2025flowdec} mainly focuses on higher bitrate settings and follows a staged design based on a waveform codec and an STFT-domain CFM post-filter, thereby motivating the study of fully integrated, lightweight architectures for extreme low-bitrate speech coding.}

\begin{figure*}
    \centering
    \includegraphics[width=0.9\linewidth]{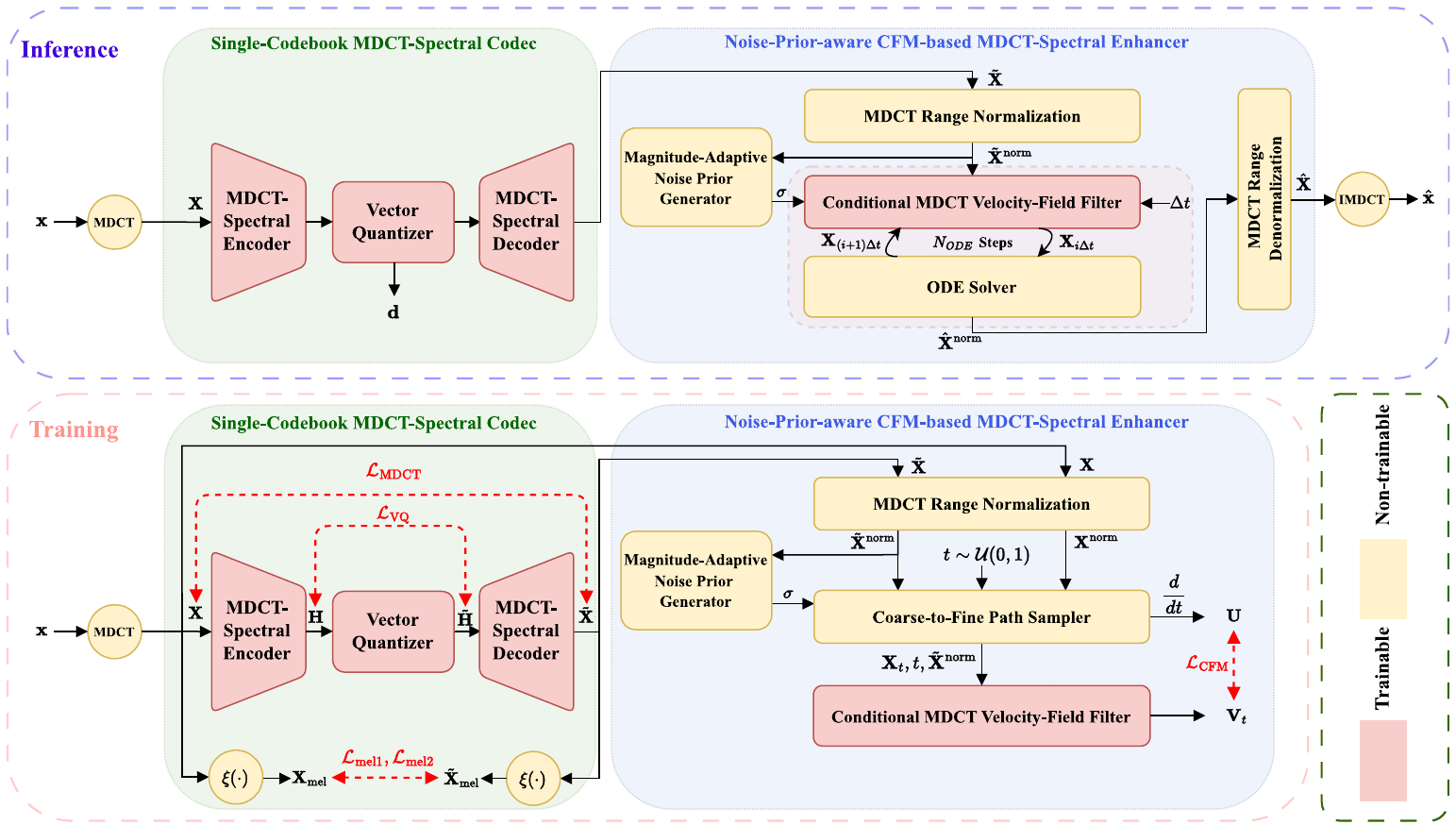}
    \caption{{An overview of the proposed CFMDCTCodec.}
    }
    \label{fig:overview}
\end{figure*}

\section{Proposed Method}
\label{sec:method}

\subsection{Overview}
\label{subsec:overview}

Fig.~\ref{fig:overview} illustrates an overview of the proposed CFMDCTCodec, which integrates a single-codebook MDCT-spectral codec with a noise-prior-aware CFM-based MDCT-spectral enhancer, with both components operating entirely in the MDCT domain.
Given an input raw speech waveform $\mathbf{x}\in\mathbb{R}^{T}$, the MDCT produces a real-valued time–frequency spectral representation $\mathbf{X}\in\mathbb{R}^{N\times K}$, where $T$ denotes the waveform length, $N$ and $K$ represent the numbers of time frames and frequency bins in the MDCT spectrum, respectively. 
With an MDCT hop size of $h_s$, the number of MDCT frames satisfies $N = \frac{T}{h_s}$.
The single-codebook MDCT-spectral codec encodes $\mathbf{X}$ into a low-rate latent sequence, applies single-codebook vector quantization for low-bitrate discretization, and decodes it to a coarse MDCT spectrum $\tilde{\mathbf{X}}\in\mathbb{R}^{N\times K}$. 
The noise-prior-aware CFM-based MDCT-spectral enhancer further improves the quality of the coarse MDCT spectrum $\tilde{\mathbf{X}}$, producing an enhanced MDCT spectrum $\hat{\mathbf{X}}\in\mathbb{R}^{N\times K}$ by leveraging CFM techniques tailored to the characteristics of MDCT representations. 
Finally, the enhanced MDCT spectrum is converted back to the speech waveform $\hat{\bm{x}}\in\mathbb R^T$ via IMDCT. 
During the training of CFMDCTCodec, we propose an end-to-end joint optimization scheme that simultaneously trains the MDCT-spectral codec and enhancer under reconstruction, quantization and CFM objectives. 
The detailed description of CFMDCTCodec is presented as follows.

\begin{figure*}[t]
  \centering
  \includegraphics[width=0.95\linewidth]{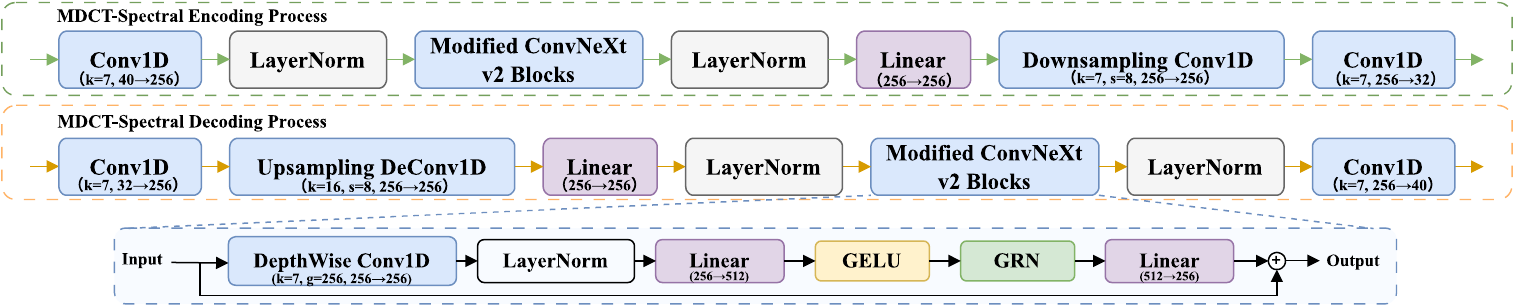}
\caption{{Architecture of the MDCT-spectral codec used in CFMDCTCodec, including the MDCT-spectral encoder and decoder. The inset at the bottom illustrates the internal structure of one modified ConvNeXt v2 block. Here, $k$ denotes the kernel size, $s$ denotes the stride, and $g$ denotes the number of convolution groups. The notation $a \rightarrow b$ indicates the change in channels.}}
  \label{fig:encoder&decoder}
\end{figure*}

\subsection{Single-Codebook MDCT-Spectral Codec}
\label{subsec:mdctcodec} 

The single-codebook MDCT-spectral codec $\phi$ draws inspiration from our previous work \cite{jiang2024mdctcodec}, yet introduces a key distinction: rather than employing RVQ, we utilize a single codebook VQ to achieve lower bitrates. 
Furthermore, we incorporate a codebook forced updating strategy during training, effectively addressing the issue of codebook collapse. 
The single-codebook MDCT-spectral codec comprises an MDCT-spectral encoder, a VQ, and an MDCT-spectral decoder. 
It discretizes the input MDCT spectrum $\bm{X}$ into a token sequence $\mathbf{d}=[d_1,\dots,d_l,\dots,d_L]^\top$, which is subsequently decoded into a coarse MDCT spectrum $\tilde{\mathbf{X}}$, where $L$ represents the length of the token sequence.
    
\subsubsection{\textbf{MDCT-Spectral Encoder \& Decoder}}
The MDCT-spectral encoder transforms the input MDCT spectrum $\mathbf{X}\in\mathbb R^{N\times K}$ into a more compact latent feature $\mathbf{H}=[\mathbf{h}_1,\dots,\mathbf{h}_l,\dots,\mathbf{h}_L]^\top\in\mathbb R^{L\times C_q}$, where $\mathbf{h}_l\in\mathbb{R}^{C_q}$ and $L< N$.
As shown in Fig.~\ref{fig:encoder&decoder}, the MDCT-spectral encoder begins with a 1-D convolutional front, followed by layer normalization. 
The output is then processed through a modified ConvNeXt-v2-style backbone \cite{woo2023convnext} for deep feature processing. 
After this, the features undergo another layer normalization operation and a linear transformation. 
To further compress the temporal resolution, the features are downsampled by a factor of $R$ using strided 1-D convolution (i.e., $R=\frac{N}{L}$). 
Finally, a 1-D convolutional backend is applied to adjust the dimensions and produce the final output. 
The modified ConvNeXt-v2-style backbone plays a crucial role in feature processing. 
It is composed of eight modified ConvNeXt-v2 blocks stacked sequentially, with each block utilizing a residual connection structure. 
Each block consists of a 1-D depthwise convolution, followed by layer normalization, pointwise linear layers, global response normalization (GRN), and Gaussian error linear unit (GELU) activation. 
As shown in Fig.~\ref{fig:encoder&decoder}, the MDCT-spectral decoder mirrors the encoder by replacing downsampling with upsampling and decodes a coarse MDCT spectrum $\tilde{\mathbf{X}}\in\mathbb{R}^{N \times K}$ from the quantized result $\tilde{\mathbf{H}}=[\tilde{\mathbf{h}}_1,\dots,\tilde{\mathbf{h}}_l,\dots,\tilde{\mathbf{h}}_L]^\top\in\mathbb R^{L\times C_q}$ of the VQ, where $\tilde{\mathbf{h}}_l\in\mathbb{R}^{C_q}$.

\begin{figure*}[t]
  \centering
  \includegraphics[width=\linewidth]{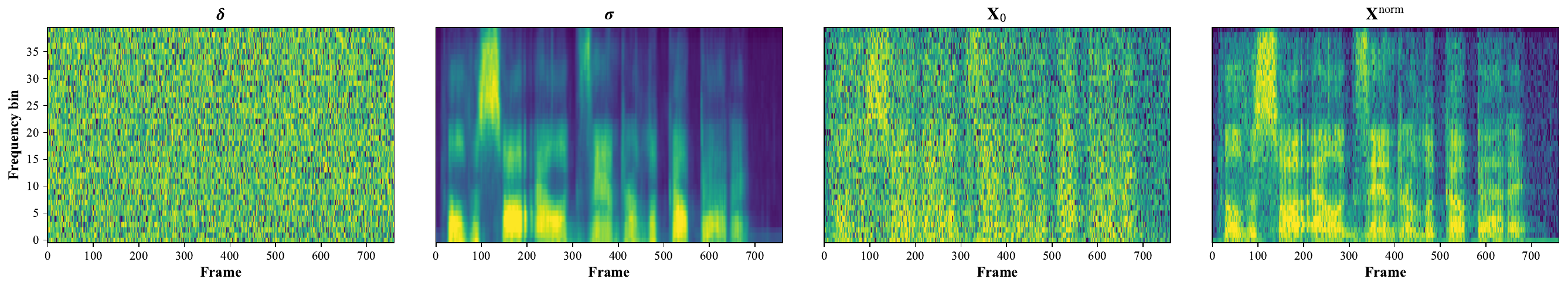}
\caption{
Visualization of the Gaussian noise $\boldsymbol{\delta}$, magnitude-adaptive noise prior $\boldsymbol{\sigma}$, CFM initial state $\mathbf{X}_0$ and CFM target terminal state $\mathbf{X}^{\mathrm{norm}}$.
}
  \label{fig:sigma_heatmap}
\end{figure*}
\subsubsection{\textbf{Single-Codebook Vector Quantization}}
The single-codebook VQ discretizes the latent feature $\mathbf{H}$ output by the MDCT-spectral encoder, generating the discrete token sequence $\mathbf{d}$ and producing the quantized result $\tilde{\mathbf{H}}$, with a trainable codebook ${\mathcal{E}}=\{\mathbf{e}_m\in\mathbb{R}^{C_q}\mid m=1,\dots,|{\mathcal{E}}|\}$, where $|{\mathcal{E}}|$ denotes the codebook size.
{During discretization, each latent vector is assigned to its nearest codevector in Euclidean distance.}
For example, consider $\mathbf{h}_l$, where $l=1,\dots,L$; its quantization process is as follows:
\begin{equation}
d_l
= \arg\min_{m\in\{1,\dots,|\mathcal{E}|\}}
\Bigl\|\tfrac{\mathbf{h}_l}{\|\mathbf{h}_l\|_2}
-
\tfrac{\mathbf{e}_m}{\|\mathbf{e}_m\|_2}\Bigr\|_2,
\end{equation}
\begin{equation}
\hat{\mathbf{h}}_l=\mathbf{e}_{d_l}.
\end{equation}

The bitrate of the discrete tokens is calculated as follows:
\begin{equation}
\label{eq:bitrate}
\mathrm{Bitrate}=\frac{f_s}{h_s\cdot R}\log_2 |\mathcal{E}| \;\; \text{(bps)},
\end{equation}
where $f_s$ (Hz) is the sampling rate of the waveform $\mathbf{x}$. 

While the single-codebook MDCT-spectral codec facilitates low-bitrate compression, the simplistic quantization unavoidably leads to the loss of fine spectral details, resulting in suboptimal speech reconstruction. 
Consequently, the decoded coarse MDCT spectrum undergoes further enhancement through a noise-prior-aware CFM-based MDCT enhancer, effectively improving the spectral quality without any increase in bitrate.

\subsection{Noise-Prior-aware CFM-based MDCT-Spectral Enhancer}
\label{subsec:fmdecoder}

The noise-prior-aware CFM-based MDCT-spectral enhancer is designed to further enhance the quality of the decoded coarse MDCT spectrum $\tilde{\mathbf{X}} \in \mathbb{R}^{N \times K}$ and generate the enhanced result $\hat{\mathbf{X}} \in \mathbb{R}^{N \times K}$. 
Specifically, the process begins by normalizing the MDCT spectrum to a well-scaled range, which helps ensure numerical stability and facilitates more effective processing. 
Based on the normalized MDCT spectrum, a magnitude-adaptive noise prior is constructed, allowing for more focused exploration in regions with higher energy. 
All subsequent CFM operations are performed in this normalized MDCT space.
The enhancement is achieved using a conditional MDCT velocity-field filter and an ODE solver. 
Finally, the solution is denormalized to produce the ultimate enhanced MDCT spectrum.

\subsubsection{\textbf{MDCT-Spectral Range Normalization/Denormalization}}
\label{subsubsec:mdct_norm}

Unlike log-amplitude STFT representations, which are strictly nonnegative, the MDCT produces a real-valued spectrum with both positive and negative coefficients, as MDCT coefficients correspond to the projection of the speech signal onto a cosine basis function. 
In practice, MDCT spectra exhibit heavy-tailed distributions and are dependent on the utterance: most coefficients are near zero, while a small subset, particularly those near formants or strong harmonics, can be several orders of magnitude larger in absolute value.
The combination of signed coefficients and highly non-uniform magnitudes makes stable CFM modeling in the raw MDCT space more challenging.

Therefore, in the noise-prior-aware CFM-based MDCT-spectral enhancer, we first normalize the range of the coarse MDCT spectrum $\tilde{\mathbf{X}}$, after which it is passed through the conditional MDCT velocity-field filter within the CFM mechanism.
MDCT range normalization addresses the dynamic range issue by applying a reversible nonlinear compression to the \emph{magnitude} while preserving the \emph{sign} within each utterance. 
Assume that $\tilde{X}_{n,k}$ is the MDCT coefficient value at the $k$-th frequency bin of the $n$-th frame of $\tilde{\mathbf{X}}$ for a given utterance. 
We first apply a power-law compression with exponent $\alpha \in (0,1]$ to the magnitudes and normalize them by the maximum compressed magnitude over the utterance, i.e.,
\begin{equation}
    |\tilde{X}|_{\max} = \max_{n,k} |\tilde{X}_{n,k}|^{\alpha}, 
\end{equation}
\begin{equation}
    \tilde{X}^{\mathrm{norm}}_{n,k}
    = \mathrm{sign}(\tilde{X}_{n,k})
      \frac{|\tilde{X}_{n,k}|^{\alpha}}{|\tilde{X}|_{\max}}.
\end{equation}
The resulting normalized values $\tilde{X}^{\mathrm{norm}}_{n,k} \in [-1, 1]$ are obtained by traversing over $n$ and $k$, yielding the normalized MDCT spectrum $\tilde{\mathbf{X}}^{\mathrm{norm}} \in \mathbb{R}^{N \times K}$. 
The pair $(\tilde{\mathbf{X}}^{\mathrm{norm}}, |\tilde{X}|_{\max})$ is stored and later used for denormalization.

As the subsequent CFM operations are carried out in the normalized MDCT-spectral domain, the enhanced normalized MDCT spectrum, $\hat{\mathbf{X}}^{\mathrm{norm}}\in \mathbb{R}^{N \times K}$, must undergo denormalization. 
The MDCT range normalization described above is fully reversible.
Assume that $\hat{X}^{\mathrm{norm}}_{n,k}$ is the element of $\hat{\mathbf{X}}^{\mathrm{norm}}$, it is denormalized by combining it with the corresponding scale $|\tilde{X}|_{\max}$ as follows, i.e.,
\begin{equation}
    \hat{X}_{n,k}
    = \mathrm{sign}(\hat{X}^{\mathrm{norm}}_{n,k})
      \bigl(|\tilde{X}|_{\max} \, \cdot|\hat{X}^{\mathrm{norm}}_{n,k}|\bigr)^{\frac{1}{\alpha}},
\end{equation}
to construct the enhanced MDCT spectrum $\hat{\mathbf{X}} \in \mathbb{R}^{N \times K}$. 
In practice, we use a small exponent, which results in more aggressive compression of large-magnitude MDCT coefficients compared to smaller ones. 
This enhances the numerical stability of the CFM objective and reduces the sensitivity of the learned flow to loudness variations at the utterance level.

\begin{figure*}[t]
  \centering
  \includegraphics[width=0.9\linewidth]{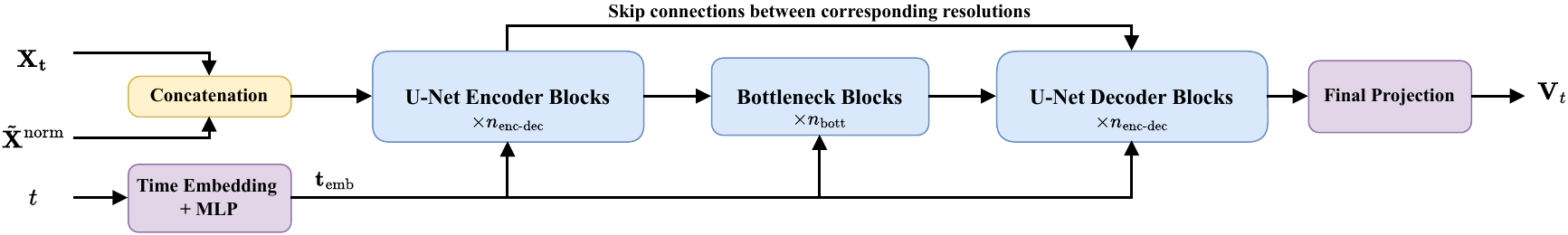}
  \caption{Details of structures of the conditional MDCT velocity-field filter.
  }
  \label{fig:velocitynet}
\end{figure*}

\subsubsection{\textbf{Magnitude-Adaptive Noise-Prior Generation}}
\label{subsubsec:energy_sigma}
Unlike the conventional approach, where CFM evolves from Gaussian noise, which can complicate the process, we use a magnitude-adaptive noise prior derived from the normalized coarse MDCT spectrum to construct the initial state of the CFM in the noise-prior-aware CFM-based MDCT-spectral enhancer. 
This provides a more informed initialization and prior, easing the CFM evolution and enhancing its efficiency.

Specifically, given the normalized coarse MDCT spectrum $\tilde{\mathbf{X}}^{\mathrm{norm}}$, we construct a magnitude-adaptive noise prior $\boldsymbol{\sigma} \in \mathbb{R}^{N \times K}$ element-wise.
We begin by computing the raw magnitude map $\mathbf{M} = |\tilde{\mathbf{X}}^{\mathrm{norm}}|$. To capture the local spectral envelope and suppress outliers, we apply 2-D average pooling to $\mathbf{M}$ with a kernel size of $(k_\text{T}, k_\text{F})$, unit stride, and same padding along both the time and frequency dimensions. 
This results in a smoothed map $\bar{\mathbf{M}} \in \mathbb{R}^{N \times K}$ that retains the original resolution.
{To reduce the influence of isolated high-energy peaks on the noise-prior scale distribution,} we then compress the smoothed magnitude map to obtain $\mathbf{M}^{\text{comp}} = \sqrt{\bar{\mathbf{M}} + \epsilon}$, where $\epsilon$ is a small constant added for numerical stability. 
The square-root operation serves as a concave dynamic-range compression, mitigating the influence of sporadic high-energy peaks on the scale distribution and resulting in a more robust variance field. 
To ensure robustness across utterances with varying loudness, we normalize $\mathbf{M}^{\text{comp}}$ using a per-utterance reference $\eta$, defined as the 99-th percentile of the magnitudes in the current utterance. 
The normalization by $\eta$ adapts to the overall loudness of the input on a global scale. 
The element $\sigma_{n,k}$ of the noise prior matrix $\boldsymbol{\sigma}$ is then derived by clipping the normalized magnitudes, i.e.,
\begin{equation}
\sigma_{n,k}=
\begin{cases}
\sigma_{\min}, & \text{if } \dfrac{M_{n,k}^{\text{comp}}}{\eta} < \sigma_{\min},\\[6pt]
\dfrac{M_{n,k}^{\text{comp}}}{\eta}, & \text{if } \sigma_{\min} \le \dfrac{M_{n,k}^{\text{comp}}}{\eta} \le \sigma_{\max},\\[6pt]
\sigma_{\max}, & \text{if } \dfrac{M_{n,k}^{\text{comp}}}{\eta} > \sigma_{\max}.
\end{cases}
\end{equation}
where $M_{n,k}^{\text{comp}}$ is the element in $\mathbf{M}^{\text{comp}}$, and $[\sigma_{\min}, \sigma_{\max}]$ defines the allowable noise range. 
Therefore, the noise prior is inherently magnitude-adaptive, being correlated with the magnitude of the MDCT spectrum.

Finally, we initialize the CFM state $\mathbf{X}_0\in \mathbb{R}^{N \times K}$ according to the noise prior $\boldsymbol{\sigma}$. 
Specifically, to ensure that $\mathbf{X}_0$ incorporates prior knowledge while preserving an element of randomness, we use $\boldsymbol{\sigma}$ as a scaling factor for Gaussian noise $\boldsymbol{\delta} \sim \mathcal{N}(\mathbf{0}, \mathbf{I})$.
Additionally, we incorporate the normalized MDCT spectrum $\tilde{\mathbf{X}}^{\mathrm{norm}}$ to further strengthen the prior knowledge, i.e.,
\begin{equation}
\label{equ8}
    \mathbf{X}_0
    = \tilde{\mathbf{X}}^{\mathrm{norm}}
      + \tau \, \boldsymbol{\sigma} \odot \boldsymbol{\delta},
\end{equation}
where $\tau$ is a temperature parameter and $\odot$ represents the element-wise multiplication. 

To provide a more intuitive demonstration of the role of the noise prior, we visualize the Gaussian noise $\boldsymbol{\delta}$, the manually constructed noise prior $\boldsymbol{\sigma}$ and CFM initial state $\mathbf{X}_0$, and the target terminal state of CFM, i.e., $\mathbf{X}^{\text{norm}}$, in Fig. \ref{fig:sigma_heatmap} for an example utterance. 
We can see that the noise prior $\boldsymbol{\sigma}$ effectively distinguishes between energetic and low-energy regions. 
When it is used as a scaling factor for Gaussian noise to construct $\mathbf{X}_0$, energetic regions are assigned heavier noise, while low-energy regions remain close to the reliable coarse MDCT spectrum. 
This is because high-energy regions of the coarse MDCT spectrum are more prone to distortion and require stronger noise for effective exploration, while low-energy regions are less affected and can be initialized closer to the original values. 
We can also see from Fig. \ref{fig:sigma_heatmap} that using $\mathbf{X}_0$ as the initial state for CFM, rather than Gaussian noise $\boldsymbol{\delta}$ as is commonly done, has the advantage that $\mathbf{X}_0$ already contains the basic shape of the MDCT spectrum. 
This makes it easier for the evolution to reach $\mathbf{X}^{\text{norm}}$, thus simplifying the CFM process.

\subsubsection{\textbf{Conditional MDCT Velocity-Field Filter}}
\label{subsubsec:velocitynet}

The CFM mechanism is designed to evolve the initial state $\mathbf{X}_0$ along the flow time axis toward the target terminal state $\mathbf{X}^{\text{norm}}$, conditioned on the normalized coarse MDCT spectrum $\tilde{\mathbf{X}}^{\mathrm{norm}}$. 
Let the state at time $t \in [0,1]$ be denoted as $\mathbf{X}_t \in \mathbb{R}^{N \times K}$, with its time derivative defined as the velocity field $\mathbf{V}_t \in \mathbb{R}^{N \times K}$, i.e.,
\begin{equation}
\label{equ9}
    \frac{d\mathbf{X}_t}{dt}
    = \mathbf{V}_t,
    \quad t\in[0,1].
\end{equation}

The core of the CFM mechanism lies in the prediction of the velocity field, which drives the iterative computation of the state $\mathbf{X}_t$. 
As illustrated in Fig.~\ref{fig:velocitynet}, we use a conditional MDCT velocity-field filter $\theta$ to estimate the velocity field. 
It adopts a lightweight 1-D U-Net architecture inspired by \cite{mehta2024matcha}. 
At each flow time $t$, the input is constructed by concatenating the state $\mathbf{X}_t$ and the conditioned $\tilde{\mathbf{X}}^{\mathrm{norm}}$ along the frequency axis. 
The scalar time $t$ is also encoded using a sinusoidal positional encoder, followed by a small multi-layer perceptron (MLP) to generate a time embedding $\mathbf{t}_{\mathrm{emb}}$. 
This embedding is then injected into all U-Net blocks to modulate feature processing throughout the flow.
After the inputs pass through the filtering network, the velocity field is generated. 
Consequently, the velocity field can be expressed as
\begin{equation}
    \mathbf{V}_t=\mathbf{V}_{\theta}(\mathbf{X}_t, t, \tilde{\mathbf{X}}^{\mathrm{norm}}).
\end{equation}

The filtering network follows a U-Net encoder--bottleneck--decoder structure. 
The U-Net encoder is composed of $n_{\text{enc-dec}}$ resolution stages that progressively downsample the temporal sequence. 
Each stage employs time-conditioned convolutional residual processing, along with lightweight temporal modeling blocks such as Transformers, to capture both local context and long-range dependencies. 
The U-Net bottleneck module contains $n_{\text{bott}}$ time-conditioned blocks at the lowest resolution. 
The U-Net decoder mirrors the encoder, consisting of $n_{\text{enc-dec}}$ stages that progressively upsample the sequence. 
Importantly, we employ skip connections between corresponding resolutions. 
The U-Net encoder feature map at each resolution is cached and, after upsampling, fused into the U-Net decoder at the same resolution through channel-wise concatenation. 
This integration provides fine-grained details that support accurate enhancement. 
Finally, a $1 \times 1$ convolution projects the U-Net decoder output back to $K$ channels, producing the predicted velocity field $\mathbf{V}_t$.

\subsubsection{\textbf{ODE Solver for Terminal State Prediction}}
\label{subsubsec:inference}

Given the trained conditional MDCT velocity-field filter, which can generate the velocity field at any given time, we use the ODE solver to iteratively predict the terminal state $\mathbf{X}_1$ from the initial state $\mathbf{X}_0$. 
Specifically, we use an explicit {Euler} solver with $N_{\mathrm{ODE}}$ uniform steps and a step size of $\Delta t = 1/N_{\mathrm{ODE}}$, iteratively executing the following equation from $i = 0$ to $i=N_{\mathrm{ODE}}-1$:
\begin{equation}
\mathbf{X}_{(i+1)\Delta t}=\mathbf{X}_{i\Delta t}+\mathbf{V}_{i\Delta t}\cdot\Delta t.
\end{equation}
The terminal state, $\mathbf{X}_1$, corresponds to the enhanced normalized MDCT spectrum, $\hat{\mathbf{X}}^{\mathrm{norm}}$, which is subsequently processed through MDCT range denormalization to yield the final enhanced MDCT spectrum $\hat{\mathbf{X}}$.

\subsection{End-to-End Joint Training Scheme}
\label{sec:training}

As shown in Fig. \ref{fig:overview}, when training CFMDCTCodec, we employ an end-to-end joint training scheme, jointly optimizing the single-codebook MDCT-spectral codec $\phi$ and the conditional MDCT velocity-field filter $\theta$ in the noise-prior-aware CFM-based MDCT-spectral enhancer. 
The loss function used for training consists of the following components.
\subsubsection{\textbf{Spectral Reconstruction Loss}}
\label{subsec:loss_spec}

The spectral reconstruction loss aims to narrow the gap between the decoded output of the single-codebook MDCT-spectral codec $\phi$ and the ground truth in the spectral domain. 
On one hand, we directly define the L2 loss $\mathcal L_{\mathrm{MDCT}}$ between the decoded coarse MDCT spectrum $\tilde{\mathbf{X}}$ and the natural one $\mathbf{X}$, reducing the spectral discrepancy in the MDCT domain. 
On the other hand, to reduce the perceptual discrepancy between the decoded result and the ground truth, we transform $\tilde{\mathbf{X}}$ and $\mathbf{X}$ into mel-spectrograms{, denoted as $\tilde{\mathbf{X}}_{\mathrm{mel}}$ and $\mathbf{X}_{\mathrm{mel}}$,} and compute the L1 and L2 losses ($\mathcal L_{\mathrm{mel1}}$ and $\mathcal L_{\mathrm{mel2}}$) in mel domain. 
The overall spectral reconstruction loss is a linear combination of the aforementioned losses, i.e.,
\begin{equation}
\begin{aligned}
\label{eq:L_spec}
\mathcal{L}_{\mathrm{spec}}(\phi)
=
\mathbb{E}_{\tilde{\mathbf{X}},\mathbf{X}}\Big[
\lambda_{\mathrm{MDCT}}\|\tilde{\mathbf{X}}-\mathbf{X}\|_{2}^{2}
+ & \lambda_{\mathrm{mel1}}\|{\tilde{\mathbf{X}}_{\mathrm{mel}}-\mathbf{X}_{\mathrm{mel}}}\|_{1}
\\+&\lambda_{\mathrm{mel2}}\|{\tilde{\mathbf{X}}_{\mathrm{mel}}-\mathbf{X}_{\mathrm{mel}}}\|_{2}^{2}
\Big],
\end{aligned}
\end{equation}
where {$\tilde{\mathbf{X}}_{\mathrm{mel}}=\xi(\tilde{\mathbf{X}})$ and $\mathbf{X}_{\mathrm{mel}}=\xi(\mathbf{X})$, and} $\xi(\cdot)=\mathbf{fb}(|\text{STFT}(\text{IMDCT}(\cdot))|)$ refers to the process of inversely transforming the MDCT spectrum into the waveform, followed by the calculation of the mel-spectrogram. 
$\mathbf{fb}$ denotes the filterbank consisting of 80 mel filters. 
$\lambda_{\mathrm{MDCT}}$, $\lambda_{\mathrm{mel1}}$ and $\lambda_{\mathrm{mel2}}$ are loss weight hyperparameters.

\subsubsection{\textbf{Quantization Loss with Codebook Forced Updating}}
\label{subsec:loss_vq}

To reduce the quantization error of the VQ in the single-codebook MDCT-spectral codec $\phi$, we first introduce the standard quantization loss defined between the input and output of the VQ, i.e., $\mathbf{H}$ and $\tilde{\mathbf{H}}$, as follows:
\begin{equation}
\label{eq:L_vq}
\mathcal{L}_{\mathrm{VQ}}(\phi)
=
\mathbb{E}_{\tilde{\mathbf{H}},\mathbf{H}}\Big[\lambda_{\mathrm{code}}
\|\operatorname{sg}[\mathbf{H}]-\tilde{\mathbf{H}}\|_{2}^{2}
+
\lambda_{\mathrm{com}}\|\mathbf{H}-\operatorname{sg}[\tilde{\mathbf{H}}]\|_{2}^{2}
\Big],
\end{equation}
where $\operatorname{sg}[\cdot]$ is the stop-gradient operator and $\lambda_{\mathrm{code}}$ and $\lambda_{\mathrm{com}}$ are the codebook loss and commitment loss weight hyperparameters.

To improve codebook utilization under a single codebook and avoid the codebook collapse issue, we then introduce a codebook forced updating strategy during VQ training, inspired by \cite{zheng2025ervq}. 
For $m$-th codevector $\mathbf{e}_m\in\mathbb{R}^{C_q}$ in the codebook $\mathcal{E}$, where $m=1,\dots,|\mathcal{E}|$, we define its \textit{assignment probability} as $p_m$. 
The $p_m$ is updated at each training step based on the previous step's result, i.e.,
\begin{equation}
p_m \leftarrow \gamma p_m + (1-\gamma)\,\bar{u}_m,
\end{equation}
where $\gamma$ is a tuning factor. 
$\bar{u}_m$ denotes the average utilization of the codevector $\mathbf{e}_m$ within a minibatch, defined as the ratio of the number of frames that select $\mathbf{e}_m$ to the total number of frames in the minibatch. 
Then, we compute the \textit{update probability} based on $p_m$ as 
\begin{equation}
\eta_m=\exp\!\left(-\frac{10\, p_m|\mathcal{E}|\,}{1-\gamma}-\zeta\right),
\end{equation}
where $\zeta$ is a small offset. 
Finally, we update $\mathbf{e}_m$ at each training step based on the update probability $\eta_m$, using the previous step's value and a sampled anchor $\mathbf{f}_m \in \mathbb{R}^{C_q}$, i.e.,
\begin{equation}
\mathbf{e}_m \leftarrow (1-\eta_m)\,\mathbf{e}_m + \eta_m\,\mathbf{f}_m,
\end{equation} 
where $\mathbf{f}_m$ is sampled from the encoded latent feature vectors of a minibatch. 
Therefore, when the update probability is high, i.e., the codevector $\mathbf{e}_m$ is infrequently utilized, it is forcibly updated to a region closer to the encoded feature vectors, thereby increasing its probability of selection; in contrast, its update is constrained. 
This strategy effectively activates underused codevectors, enhances codebook utilization, and consequently improves coding performance.

\subsubsection{\textbf{CFM Loss}}
\label{subsec:loss_fm}

The CFM loss aims to minimize the discrepancy between the velocity field predicted by the conditional MDCT velocity-field filter $\theta$ in the noise-prior-aware CFM-based MDCT-spectral enhancer and the true value, thereby promoting accurate prediction of the CFM terminal state, i.e., $\mathbf{X}^{\text{norm}}$. 
Specifically, we assume that the evolution from the initial state $\mathbf{X}_0$ defined by Equation \ref{equ8} to the final state $\mathbf{X}^{\text{norm}}$ occurs linearly along the flow time axis, i.e.,
\begin{equation}
\mathbf{X}_t = \mathbf{X}_0 + t(\mathbf{X}^{\text{norm}}-\mathbf{X}_0),
\quad
t\in[0,1],
\end{equation}
which can also be regarded as a coarse-to-fine path sampler. 
According to equation \ref{equ9}, the target velocity field is defined as
\begin{equation}
\mathbf{U} = \mathbf{X}^{\text{norm}}-\mathbf{X}_0,
\end{equation}
which is independent of flow time. 
The CFM loss is defined as the L2 loss between the predicted velocity field $\mathbf{V}_t = \mathbf{V}_{\theta}(\mathbf{X}_t, t, \tilde{\mathbf{X}}^{\mathrm{norm}})$ and the target velocity field $\mathbf{U}$, i.e.,
\begin{equation}
\label{eq:L_fm}
\mathcal{L}_{\mathrm{CFM}}(\phi,\theta)
=
\mathbb{E}_{t\sim\mathcal{U}(0,1)}
\big[
\|\mathbf{V}_t-\mathbf{U}\|_2^2
\big].
\end{equation}
The gradient of the CFM loss is propagated through both $\phi$ and $\theta$, facilitating the concurrent optimization of these two components. 

\subsubsection{\textbf{Overall Training Loss}}
\label{subsec:loss_total}

The above losses are combined to jointly optimize the trainable components of CFMDCTCodec (i.e., $\phi$ and $\theta$):
\begin{equation}
\label{eq:L_total}
\mathcal{L}(\phi,\theta)
=
\mathcal{L}_{\mathrm{spec}}(\phi)
+
\mathcal{L}_{\mathrm{VQ}}(\phi)
+
\lambda_{\mathrm{CFM}}\,\mathcal{L}_{\mathrm{CFM}}(\phi,\theta),
\end{equation}
where $\lambda_{\mathrm{CFM}}$ is a loss weight hyperparameter.

\subsection{{Relationship to Prior Generative Speech Codecs}}
\label{subsec:relationship}

{The preceding subsections have presented the technical details of CFMDCTCodec in depth. 
Building on that foundation, this subsection provides a systematic discussion and summary of the commonalities and distinctions between CFMDCTCodec and prior generative speech codecs. 
CFMDCTCodec, together with FlowDec \cite{welker2025flowdec} and ScoreDec \cite{wu2024scoredec}, belongs to the family of generative speech codecs with post-filtering, in which a deterministic base codec first produces a coarse reconstruction and a generative post-filter subsequently refines it. 
These codecs share several core design principles: the use of power-law amplitude-compressed time--frequency representations, interpolation-based flow construction between the coarse estimate and the ground truth, informed modulation of the prior noise distribution, and the avoidance of adversarial training objectives.}

{Despite these commonalities, CFMDCTCodec differs from FlowDec and ScoreDec in several aspects that are particularly relevant to low-bitrate speech coding. 
At the framework level, both FlowDec and ScoreDec adopt a decoupled two-stage training paradigm in which the base codec is first trained independently and then frozen while the post-filter is optimized separately; in contrast, CFMDCTCodec jointly optimizes the MDCT-spectral codec and the CFM-based enhancer in an end-to-end manner, allowing the two components to co-adapt throughout training. 
In addition, from an overall perspective, CFMDCTCodec adopts a more lightweight and efficient architecture compared with FlowDec and ScoreDec. 
In terms of the codec design, FlowDec and ScoreDec rely on multi-codebook RVQ-based waveform codecs operating on raw samples, whereas CFMDCTCodec adopts a single-codebook quantizer with codebook forced updating in the real-valued MDCT domain. 
Regarding the enhancer, FlowDec performs enhancement in the complex STFT domain and constructs its noise prior based on dataset-level, frequency-dependent statistics. In contrast, CFMDCTCodec operates in the MDCT domain and derives a magnitude-adaptive prior for each utterance and each time--frequency bin, thereby enabling instance-dependent modulation that better captures the spectral energy distribution of individual inputs.}

\section{Experiments and results}
\label{sec:experiment}
\subsection{Experimental Setup}
\label{sec:experimental_setup}
We conducted experiments on two speech corpora at different sampling rates.
For 16-kHz experiments, we used LibriTTS \cite{zen2019libritts} and followed the standard split, using \textit{train-clean-100} and \textit{train-clean-360} for training, \textit{dev-clean} for validation, and \textit{test-clean} for evaluation.
For 48-kHz experiments, we used the VCTK dataset \cite{veaux2017superseded}, with 40,936 utterances for training and 2,937 utterances for testing.

For CFMDCTCodec, when extracting the MDCT spectrum from the speech waveform, the MDCT used a frame length, hop size, and frequency bin number of 80, 40, and 40, respectively (i.e., $h_s=K=40$). 
In the single-codebook MDCT-spectral codec of CFMDCTCodec, the MDCT-spectral encoder and decoder used the same configuration as in \cite{jiang2024mdctcodec}.
The VQ used a codebook of size 8192 (i.e., $|\mathcal{E}|=8192$), with codevector dimensions of 32 (i.e., $C_q=32$). 
In the noise-prior-aware CFM-based MDCT-spectral enhancer of the CFMDCTCodec, the compression exponent was set to $\alpha=0.5$ for MDCT-spectral range normalization. 
When generating the noise prior, the kernel size of the 2-D average pooling was set to $(k_\text{T}, k_\text{F})=(3,5)$, and the compression constant is set to $\epsilon=1\times10^{-8}$. 
The noise upper and lower bounds were set to $\sigma_{\max}=1.0$ and $\sigma_{\min}=1\times10^{-3}$, respectively, {and the temperature parameter $\tau$ was set to $1.0$ for 16~kHz and $1.3$ for 48~kHz.}
The conditional MDCT velocity-field filter had 2 blocks in each of its modules (i.e., $n_{\text{enc-dec}}=n_{\text{bott}}=2$), with other configurations matching those used in \cite{mehta2024matcha}. 
The {explicit Euler} ODE solver used a total of $N_{\mathrm{ODE}}=6$ steps, i.e., the step size was $\Delta t=\frac{1}{6}$.

When training CFMDCTCodec, the loss weight hyperparameters were set to $\lambda_{\mathrm{MDCT}}=250$, $\lambda_{\mathrm{mel1}}=20$, $\lambda_{\mathrm{mel2}}=10$, $\lambda_{\mathrm{code}}=10,\lambda_{\mathrm{com}}=2.5$ and $\lambda_{\mathrm{CFM}}=100$. 
In the codebook forced updating training strategy, the tuning factor was set to $\gamma=0.99$, and the offset is set to $\zeta=1\times10^{-3}$.
The training is performed with AdamW \cite{loshchilov2018decoupled} optimizer using learning rate $2\times10^{-4}$ and betas $(0.8,0.99)$,
together with an exponential learning-rate schedule with decay factor $0.999$. 
In each training step, a minibatch was composed of 48 1-second speech segments, i.e., the batch size was 48, and training continued until 1M training steps were completed.

We configured three bitrate settings: low, medium, and high. 
The low-bitrate scenario operated at only 0.65 kbps, and the experiment was conducted on the 16-kHz dataset, through setting the downsampling/upsampling rate of the MDCT-spectral encoder and decoder to $R=8$. 
The medium-bitrate scenario operated at 1.3 kbps and 1.95 kbps, with experiments conducted on the 16-kHz and 48-kHz datasets by setting $R=4$ and $R=8$, respectively. 
For the high-bitrate scenario, the bitrate was set as 3.9 kbps, with experiments conducted on the 48-kHz dataset with $R=4$. 
\subsection{Evaluation Metrics}
\label{subsec:metrics}

We employed both objective and subjective metrics to evaluate the performance of compared neural speech codecs. 

\begin{itemize}[leftmargin=*, labelsep=0.5em]

\item \textbf{Objective Metrics:}
We employed seven objective metrics for speech quality evaluation, including short-time objective intelligibility (STOI) \cite{taal2010short}, which measures speech intelligibility; scale-invariant signal-to-distortion ratio (SI-SDR) \cite{le2019sdr}, used to evaluate time-domain waveform fidelity; speaker similarity (SIM) \cite{chen2022wavlm}, which measures the preservation of speaker identity; {log-spectral distance (LSD) \cite{gray2003distance}, which measures the spectral distortion between the reconstructed and reference speech in the log-spectral domain;} DNSMOS \cite{reddy2021dnsmos}, a non-intrusive neural quality estimator predicting mean opinion scores (MOS) for 16-kHz speech; SIGMOS \cite{ristea2025icassp}, a full-band non-intrusive quality predictor for 48-kHz speech; and UTMOS \cite{saeki2022utmos}, a non-intrusive domain-robust metric predicting human subjective ratings with high correlation to human perception in synthesized speech tasks.
{Additionally, to evaluate generation efficiency, the real-time factor (RTF) was measured, defined as the ratio of the generation time to the actual duration of the audio. RTF values were computed by averaging over the entire test set on a single Intel Xeon Silver 4210 CPU  {using 10 CPU cores} and a single NVIDIA A100 GPU, respectively.} To assess the computational and model complexity of a codec, we also reported the floating-point operations (FLOPs) required for generating 1-second of speech, as well as the total number of trainable parameters (Param.).

\item \textbf{Subjective Metric:} 
We conducted multiple stimuli with hidden reference and anchor (MUSHRA) \cite{series2014method} tests, one for each of the four bitrate settings, on the crowdsourcing Amazon Mechanical Turk (AMT) platform\footnote{\href{https://www.mturk.com}{https://www.mturk.com}.} to evaluate the subjective quality of the speech generated by different codecs at equal bitrate. 
20 test utterances generated by each experimental codec were evaluated by a total of 30 English native listeners. 
{Listeners were asked to give a score between 0 and 100 to each test sample, with natural speech as the hidden reference and a 3.5-kHz low-pass-filtered version as the anchor.}

\end{itemize}
\begin{table*}[t]
\small
\centering
\caption{Objective and subjective quality-related experimental results for CFMDCTCodec and baselines on the LibriTTS test set (16 kHz) at low bitrate (0.65 kbps) and medium bitrate (1.3 kbps). {MUSHRA scores of hidden reference and anchor for 0.65 kbps were 93.95$\pm$2.24 and 40.83$\pm$5.12, respectively; MUSHRA scores of hidden reference and anchor for 1.3 kbps were 95.63$\pm$2.16 and 41.54$\pm$5.49, respectively. The \textbf{bold} and \underline{underlined} numbers indicate optimal and sub-optimal results, respectively.}}
\begin{tabular}{l|c c|c c c c c c|c}
\hline

\hline
\textbf{Codec} & \textbf{Bitrate} & \textbf{{Frame rate}}
& \textbf{STOI}$\uparrow$ & \textbf{SI-SDR}$\uparrow$ & \textbf{SIM}$\uparrow$ & {\textbf{LSD}$\downarrow$} & \textbf{DNSMOS}$\uparrow$ & \textbf{UTMOS}$\uparrow$ 
& \textbf{MUSHRA}$\uparrow$ \\
\hline
MDCTCodec    & \multirow{6}{*}{\shortstack{0.65~kbps\\(Low)}} & \multirow{6}{*}{\shortstack{{50 Hz}}} & \underline{0.870} & -16.527 & \textbf{0.945} & \textbf{{0.932}} & \underline{3.207} & 2.971 & {64.88$\pm$4.79} \\
DAC          &&                            & 0.838 & \underline{-3.578}  & 0.922 & {0.978} & 3.157 & 2.884 & {71.83$\pm$4.30} \\
BigCodec     &&                            & \textbf{0.886} & -7.787  & \underline{0.944} & \underline{{0.936}} & \textbf{3.221} & \textbf{3.846} & \textbf{{78.15$\pm$3.27}} \\
WavTokenizer &&                            & 0.829 & -20.457 & 0.911 & {0.989} & 3.090 & 3.060 & {73.80$\pm$3.82} \\
FlowDec      &&                            & 0.781 & -16.209 & 0.910 & {1.080} & 2.999 & 1.915 & {67.01$\pm$4.64} \\
CFMDCTCodec  &&                            & 0.866 & \textbf{-3.206}  & 0.942 & {1.166} & 3.186 &\underline{3.761} & \underline{{76.81$\pm$3.64}} \\
\cline{1-10}
MDCTCodec    & \multirow{6}{*}{\shortstack{1.3~kbps\\(Medium)}} & \multirow{6}{*}{\shortstack{{100 Hz}}} & \underline{0.912} & -4.254 & \underline{0.964} & \textbf{{0.884}} & \textbf{3.266} & 3.567 & {79.32$\pm$3.86} \\
DAC          &&                            & 0.894 & \textbf{1.218}  & 0.950 & {0.925} & 3.215 & 3.407 & {80.36$\pm$3.92} \\
BigCodec     &&                            & \textbf{0.922} & -2.886 & \textbf{0.969} & \underline{{0.895}} & \underline{3.235} & \textbf{4.016} & \textbf{{82.36$\pm$3.49}} \\
WavTokenizer &&                            & 0.889 & -0.106 & 0.937 & {0.949} & 3.186 & 3.788 & \underline{{82.05$\pm$3.44}} \\
FlowDec      &&                            & 0.854 & -7.252 & 0.939 & {1.009} & 3.147 & 2.794 & {69.29$\pm$4.23} \\
CFMDCTCodec  &&                            & 0.906 & \underline{0.591}  & 0.963 & {1.157} & 3.216 & \underline{3.862} & {81.66$\pm$3.65} \\
\hline

\hline
\end{tabular}
\label{tab:main_objective_16k}
\end{table*}
\begin{table*}[t]
\small
\centering
\caption{Objective and subjective quality-related experimental results for CFMDCTCodec and baselines on the test set of VCTK dataset (48 kHz) at medium bitrate (1.95 kbps) and high bitrate (3.9 kbps). {MUSHRA scores of hidden reference and anchor for 1.95 kbps were 94.26$\pm$2.15 and 41.13$\pm$5.80, respectively; MUSHRA scores of hidden reference and anchor for 3.9 kbps were 94.35$\pm$2.47 and 39.91$\pm$5.17, respectively. The \textbf{bold} and \underline{underlined} numbers indicate optimal and sub-optimal results, respectively.}}
\begin{tabular}{l|c c|c c c c c c|c}
\hline

\hline
\textbf{Codec} & \textbf{Bitrate} & \textbf{{Frame rate}}
& \textbf{STOI}$\uparrow$ & \textbf{SI-SDR}$\uparrow$ & \textbf{SIM}$\uparrow$ & \textbf{{LSD}$\downarrow$} & \textbf{SIGMOS}$\uparrow$ & \textbf{UTMOS}$\uparrow$
& \textbf{MUSHRA}$\uparrow$ \\
\hline
MDCTCodec    & \multirow{6}{*}{\shortstack{1.95~kbps\\(Medium)}} & \multirow{6}{*}{\shortstack{{150 Hz}}} & \underline{0.840} & -1.347 & \textbf{0.952} & \textbf{{0.854}} & 2.935 & \underline{3.686} & {76.02$\pm$4.10} \\
DAC          &&                            & 0.803 & \textbf{2.220}  & 0.915 & {0.884} & 2.933 & 3.344 & {74.98$\pm$4.37} \\
BigCodec     &&                            & \textbf{0.851} & \underline{1.719}  & \underline{0.941} & \underline{{0.857}} & \underline{3.213} & \textbf{3.910} & \textbf{{82.47$\pm$3.99}} \\
WavTokenizer &&                            & 0.774 & -1.876 & 0.853 & {0.961} & 2.856 & 2.987 & {66.13$\pm$5.21} \\
FlowDec      &&                            & 0.790 & -1.086 & 0.923 & {0.924} & \textbf{3.339} & 3.621 & \underline{{81.74$\pm$3.56}} \\
CFMDCTCodec  &&                            & {0.810} & {1.401} & {0.921} & {1.089} & {2.927} & {3.667} & {79.06$\pm$3.74} \\
\cline{1-10}
MDCTCodec    & \multirow{6}{*}{\shortstack{3.9~kbps\\(High)}} & \multirow{6}{*}{\shortstack{{300 Hz}}} & \underline{0.866} & 1.233 & \textbf{0.966} & \textbf{{0.838}} & 2.976 & \underline{3.808} & {82.07$\pm$3.04} \\
DAC          &&                            & 0.860 & \textbf{6.119} & 0.950 & {0.858} & 3.004 & 3.573 & {83.24$\pm$3.92} \\
BigCodec     &&                            & \textbf{0.895} & \underline{5.400} & \underline{0.958} & \underline{{0.846}} & \underline{3.230} & \textbf{3.931} & \underline{{84.35$\pm$3.12}} \\
WavTokenizer &&                            & 0.856 & 4.401 & 0.923 & {0.878} & 3.126 & 3.788 & {83.59$\pm$3.28} \\
FlowDec      &&                            & 0.825 & 0.826 & 0.951 & {0.873} & \textbf{3.363} & 3.773 & \textbf{{85.98$\pm$3.65}} \\
CFMDCTCodec  &&                            & {0.837} & {4.364} & {0.943} & {1.056} & {3.041} & {3.772} & {83.37$\pm$3.79} \\
\hline

\hline
\end{tabular}
\label{tab:main_objective_48k}
\end{table*}

\subsection{Baseline Configuration}
\label{subsec:baseline_cmp}

We compared CFMDCTCodec with several representative neural speech codecs, encompassing a range of design choices in modeling targets, quantization methods, and model capacity. 
We adjusted the configurations to match the bitrate settings we used. 
These baselines include:

\begin{itemize}[leftmargin=*, labelsep=0.5em]
  \item \textbf{MDCTCodec} \cite{jiang2024mdctcodec}: A lightweight spectral-domain neural speech codec that operates directly on the MDCT spectrum using RVQ discretization. 
  It does not have a post-processing module and requires adversarial training.

  \item \textbf{DAC}\footnote{\href{https://github.com/descriptinc/descript-audio-codec}{https://github.com/descriptinc/descript-audio-codec}.} \cite{kumar2024high}: A waveform-based neural speech codec with a deep encoder–decoder backbone and RVQ discretization. 
  It also relies on adversarial training and serves as a widely used baseline.

  \item \textbf{BigCodec}\footnote{\href{https://github.com/Aria-K-Alethia/BigCodec}{https://github.com/Aria-K-Alethia/BigCodec}.} \cite{xin2024bigcodec}: A waveform-based neural speech codec with a huge encoder–decoder backbone and single-codebook VQ discretization. 
  It scales the encoder and decoder to {over a hundred million} parameters to preserve speech details at the cost of very high complexity, and adopts adversarial training.

  \item \textbf{WavTokenizer}\footnote{\href{https://github.com/jishengpeng/WavTokenizer}  {https://github.com/jishengpeng/WavTokenizer}.} \cite{ji2024wavtokenizer}: A neural speech codec designed for low-bitrate scenario, which encodes waveforms but decodes STFT spectra, with single-codebook VQ discretization. 
  It enhances the reconstruction capability of the decoder and also employs adversarial training, serving as a representative work for low-bitrate coding.

  \item \textbf{FlowDec}\footnote{\href{https://github.com/facebookresearch/FlowDec}  {https://github.com/facebookresearch/FlowDec}.} \cite{welker2025flowdec}: A neural speech codec that incorporates a CFM-based postprocessor. 
  It uses an RVQ-based codec module to discretize the speech waveform and then enhances the decoded speech's STFT spectra through a postprocessor. 
  It does not employ adversarial training strategy but requires a two-stage training process, with the codec module and postprocessor trained in sequence.
\end{itemize}
{To ensure a fair comparison, we configured all baseline models to operate at the same bitrates. Following the bitrate formulation in Eq.~(\ref{eq:bitrate}), we achieved this by fixing the single codebook size to $|\mathcal{E}|=8192$ (13 bits per code) and adapting the overall temporal downsampling factor (i.e., $D=h_s\cdot R$). Specifically, for the 0.65-kbps and 1.95-kbps settings, we adjusted the strides of the downsampling convolutional layers in the waveform-based baselines (i.e., DAC, BigCodec, WavTokenizer and FlowDec) to $(2, 4, 5, 8)$, yielding a total downsampling factor of $D=320$. For the MDCT-based codecs (i.e., MDCTCodec and CFMDCTCodec), the total downsampling factor is the product of the MDCT hop size and the convolutional downsampling rate. By setting the hop size to $h_s=40$ and the downsampling rate to $R=8$, we matched the identical total downsampling factor of $D=320$.
Similarly, for the 1.3-kbps and 3.9-kbps settings, the convolutional strides for the waveform-based baselines were modified to $(2, 4, 4, 5)$ to achieve a total downsampling factor of $D=160$. For the MDCT-based codecs, the MDCT hop size remained fixed at $h_s=40$, while the convolutional downsampling rate was reduced to $R=4$.}
 {With the codebook size fixed, these stride adjustments align the output frame rate across baselines at each bitrate setting. This avoids controlling bitrate solely by changing the number of bits per code, which would require either overly small codebooks at low bitrates or impractically large single codebooks at high bitrates.}

\begin{table*}[t]
\small
\centering
\caption{{Efficiency and complexity performance comparison for CFMDCTCodec and baselines on the test set of LibriTTS dataset (16 kHz) at low bitrate (0.65 kbps) and medium bitrate (1.3 kbps), respectively. Here, ``$a\times$" represents $a\times$ real time. The \textbf{bold} and \underline{underlined} numbers indicate optimal and sub-optimal results, respectively.}}
\begin{tabular}{l|c c|c c|c c}
\hline
\hline
\textbf{Codec} & \textbf{Bitrate} & {\textbf{Frame rate}}
& {\textbf{RTF(CPU)}$\downarrow$} & {\textbf{RTF(GPU)}$\downarrow$}
& \textbf{FLOPs}$\downarrow$ & \textbf{Param.}$\downarrow$ \\
\hline
MDCTCodec    & \multirow{6}{*}{\shortstack{0.65~kbps\\(Low)}}    & \multirow{6}{*}{\shortstack{{50 Hz}}}   
& {\textbf{0.055~(18.182$\times$)}} & {\textbf{0.013~(76.923$\times$)}} & \textbf{2.28G}    & \textbf{6.15M}  \\
DAC          &                                                   &                                       
& {0.722~(1.385$\times$)}  & {0.080~(12.500$\times$)} & 55.53G   & 73.86M \\
BigCodec     &                                                   &                                       
& {1.899~(0.527$\times$)}  & {0.044~(22.727$\times$)} & {52.56G}   & {112.36M} \\
WavTokenizer &                                                   &                                       
& {\underline{0.184~(5.435$\times$)}}  & {\underline{0.020~(50.000$\times$)}} & \underline{4.20G}    & 71.65M \\
FlowDec      &                                                   &                                       
& {21.349~(0.047$\times$)} & {0.150~(6.667$\times$)}  & 2306.48G & 97.54M \\
CFMDCTCodec  &                                                   &                                       
& {0.442~(2.262$\times$)}  & {0.088~(11.364$\times$)} & 11.93G   & \underline{14.61M} \\
\cline{1-7}
MDCTCodec    & \multirow{6}{*}{\shortstack{1.3~kbps\\(Medium)}}  & \multirow{6}{*}{\shortstack{{100 Hz}}} 
& {\textbf{0.058~(17.241$\times$)}} & {\textbf{0.011~(90.909$\times$)}} & \textbf{2.30G}    & \textbf{6.15M}  \\
DAC          &                                                   &                                       
& {0.680~(1.471$\times$)}  & {0.093~(10.753$\times$)}  & 55.49G   & 62.78M \\
BigCodec     &                                                   &                                       
& {1.819~(0.550$\times$)}  & {0.051~(19.608$\times$)}  & {54.59G}   & {102.78M} \\
WavTokenizer &                                                   &                                       
& {\underline{0.326~(3.067$\times$)}}  & {\underline{0.023~(43.478$\times$)}}  & \underline{7.69G}    & 70.79M \\
FlowDec      &                                                   &                                       
& {20.063~(0.050$\times$)} & {0.151~(6.623$\times$)}   & 2306.44G & 86.46M \\
CFMDCTCodec  &                                                   &                                       
& {0.462~(2.165$\times$)}  & {0.069~(14.493$\times$)}  & 11.95G   & \underline{14.61M} \\
\hline
\hline
\end{tabular}
\label{tab:main_rtf_16k}
\end{table*}
\begin{table*}[t]
\small
\centering
\caption{{Efficiency and complexity performance comparison for CFMDCTCodec and baselines on the test set of VCTK dataset (48 kHz) at medium bitrate (1.95 kbps) and high bitrate (3.9 kbps), respectively. Here, ``$a\times$" represents $a\times$ real time. The \textbf{bold} and \underline{underlined} numbers indicate optimal and sub-optimal results, respectively.}}
\begin{tabular}{l|c c|c c|c c}
\hline
\hline
\textbf{Codec} & \textbf{Bitrate} & {\textbf{Frame rate}}
& {\textbf{RTF(CPU)}$\downarrow$} & {\textbf{RTF(GPU)}$\downarrow$}
& \textbf{FLOPs}$\downarrow$ & \textbf{Param.}$\downarrow$ \\
\hline
MDCTCodec    & \multirow{6}{*}{\shortstack{1.95~kbps\\(Medium)}} & \multirow{6}{*}{\shortstack{{150 Hz}}} & {\textbf{0.142~(7.042$\times$)}}  & {\textbf{0.019~(52.632$\times$)}} & \textbf{6.80G}    & \textbf{6.15M}  \\
DAC          &                                                    &                                       & {2.238~(0.447$\times$)}  & {0.122~(8.197$\times$)}  & 166.61G  & 73.86M \\
BigCodec     &                                                    &                                       & {6.386~(0.157$\times$)}  & {0.074~(13.514$\times$)} & {157.69G}   & {112.36M} \\
WavTokenizer &                                                    &                                       & {\underline{0.564~(1.773$\times$)}}  & {\underline{0.026~(38.462$\times$)}} & \underline{12.61G}   & 71.65M \\
FlowDec      &                                                    &                                       & {61.028~(0.016$\times$)} & {0.359~(2.786$\times$)}  & 4668.51G & 97.54M \\
CFMDCTCodec  &                                                    &                                       & {1.378~(0.726$\times$)}  & {0.124~(8.065$\times$)}  & 35.61G   & \underline{14.61M} \\
\cline{1-7}
MDCTCodec    & \multirow{6}{*}{\shortstack{3.9~kbps\\(High)}}    & \multirow{6}{*}{\shortstack{{300 Hz}}} & {\textbf{0.156~(6.410$\times$)}}  & {\textbf{0.018~(55.556$\times$)}} & \textbf{6.87G}    & \textbf{6.15M}  \\
DAC          &                                                    &                                       & {2.155~(0.464$\times$)}  & {0.159~(6.289$\times$)}  & 166.62G  & 62.78M \\
BigCodec     &                                                    &                                       & {7.214~(0.139$\times$)}  & {0.101~(9.901$\times$)}  & {163.78G}  & {102.78M} \\
WavTokenizer &                                                    &                                       & {\underline{0.973~(1.028$\times$)}}  & {\underline{0.045~(22.222$\times$)}} & \underline{23.07G}   & 70.79M \\
FlowDec      &                                                    &                                       & {58.365~(0.017$\times$)} & {0.361~(2.770$\times$)}  & 4668.52G & 86.46M \\
CFMDCTCodec  &                                                    &                                       & {1.440~(0.694$\times$)}  & {0.122~(8.197$\times$)}  & 35.68G   & \underline{14.61M} \\
\hline
\hline
\end{tabular}
\label{tab:main_rtf_48k}
\end{table*}

\subsection{Quality Comparison Analysis}
\label{subsec:baseline_cmp_results}

The objective and subjective quality-related experimental results are summarized in Tables~\ref{tab:main_objective_16k} and \ref{tab:main_objective_48k}. Since the relative performance of different codecs varies across bitrate regimes, we organize the discussion by bitrate setting below.

\subsubsection{\textbf{Low-Bitrate (0.65~kbps) Comparison}}
This paper focuses on low-bitrate coding, so we primarily analyze the performance of various codecs at 0.65 kbps on the 16-kHz LibriTTS dataset.
We first compared CFMDCTCodec with two representative codecs based on spectrum and waveform, i.e., MDCTCodec and DAC.
Compared to MDCTCodec, CFMDCTCodec exhibited a  {clear} advantage in both SI-SDR and UTMOS, while performing comparably on other quality-related objective metrics{, except for LSD}.
In terms of subjective listening quality, CFMDCTCodec’s MUSHRA score exceeded that of MDCTCodec by {approximately 12 points, indicating a noticeable perceptual improvement brought by the CFM-based enhancer  {at this extremely low bitrate}.}
The standard decoder of MDCTCodec is unable to accurately recover speech from the severely compressed bitstream.
In contrast, our CFMDCTCodec greatly enhanced decoding performance by integrating the MDCT-spectral enhancer with CFM methodology.
{Interestingly, although CFMDCTCodec achieved better subjective quality, this advantage was not reflected in LSD, where it performed worse than MDCTCodec.
This may be attributed to the fact that MDCTCodec directly optimizes spectral reconstruction, whereas CFMDCTCodec prioritizes perceptual refinement over strict spectral accuracy.
The similarly elevated LSD of FlowDec, another generative post-filtering codec, further supports this interpretation.}
Compared to the waveform-based DAC, our CFMDCTCodec substantially outperformed DAC across {most} quality-related objective and subjective metrics.

Next, we compared CFMDCTCodec with BigCodec and WavTokenizer, both of which are designed for low-bitrate scenarios and use single-codebook quantization.
As illustrated in Table~\ref{tab:main_objective_16k}, BigCodec exhibited competitive performance, achieving comparable results to our CFMDCTCodec across both quality-related objective and subjective metrics.
 {In contrast, under this condition, WavTokenizer lagged behind CFMDCTCodec on most metrics.}

Finally, we compared CFMDCTCodec with FlowDec, which also employs CFM-based postprocessing.
As reported in Table~\ref{tab:main_objective_16k}, FlowDec performed poorly in both intelligibility and perceptual quality, yielding the lowest STOI and UTMOS among the compared systems, and its subjective MUSHRA score was only marginally higher than that of MDCTCodec.
This may be attributed to the fact that FlowDec’s postprocessor was originally developed for higher-bitrate, DAC-style codecs; under such an extreme low-bitrate constraint, the conditioning becomes too degraded for the postprocessor to refine effectively.
Moreover, FlowDec’s two-stage training paradigm may be ill-suited to the low-bitrate regime, which further supports the advantage of our end-to-end joint training strategy in CFMDCTCodec.
We will provide a more detailed experimental discussion of training paradigms in Section~\ref{subsec: Training Scheme Discussion}.

\subsubsection{\textbf{Medium-Bitrate (1.3~kbps and 1.95~kbps) Comparison}}
As the bitrate increased, the perceptual gaps across codecs narrowed and most  {systems} achieved consistently strong subjective quality, suggesting that the additional bitrate provided richer conditioning information and reduced reconstruction ambiguity.
 {At 1.3~kbps, CFMDCTCodec and MDCTCodec achieved comparable MUSHRA scores, in contrast to the clear gap observed at 0.65~kbps.}
 {At this bitrate, WavTokenizer achieved MUSHRA scores on par with CFMDCTCodec, while at 1.95~kbps, WavTokenizer exhibited notably poor performance, which may be attributed to its architecture being less suited to the 48-kHz sampling rate or the specific downsampling configuration.}
 {BigCodec consistently ranked among the top performers across both medium-bitrate settings, though at the cost of significantly higher model complexity.}
 {FlowDec showed a clear bitrate-dependent trend: it underperformed at 1.3~kbps but became increasingly competitive at 1.95~kbps, suggesting that its STFT-domain postprocessor benefits more from higher-quality conditioning signals.}

\subsubsection{\textbf{High-Bitrate (3.9~kbps) Comparison}}
 {At 3.9~kbps, the perceptual gaps further narrowed and all codecs achieved good reconstruction quality.}
 {CFMDCTCodec was comparable to MDCTCodec, DAC, and WavTokenizer, and slightly below BigCodec and FlowDec.}
 {Under this less constrained setting, the advantage of the MDCT-spectral enhancer in CFMDCTCodec diminished, since the base encoder--decoder already preserved sufficient spectral detail.}

Overall, CFMDCTCodec’s main advantage emerged in the low-bitrate regime, where it delivered the most substantial quality gains.
At medium and high bitrates, it still maintained competitive performance against other strong neural speech codecs, indicating that its low-bitrate strength did not compromise performance at higher bitrates.
By comparing Tables~\ref{tab:main_objective_16k} and \ref{tab:main_objective_48k}, we can also observe a dataset effect: performance gaps were generally smaller on VCTK, which is smaller and acoustically cleaner than LibriTTS. This suggests that CFMDCTCodec preserved stronger robustness under more challenging and noisier conditions.

\subsection{Efficiency and Complexity Comparison Analysis}
\label{subsec:efficiency_complexity}

 {The efficiency and complexity results are summarized in Tables~\ref{tab:main_rtf_16k} and \ref{tab:main_rtf_48k}. We focus the discussion on the 0.65~kbps setting, as the relative trends were consistent across bitrates.}
{Although the extra enhancer came at the cost of reduced efficiency and increased complexity,  {at the 0.65~kbps setting,} CFMDCTCodec still achieved real-time generation on both CPU and GPU, and the trade-off between quality and efficiency/complexity remained acceptable.}
Compared to the waveform-based DAC, our CFMDCTCodec demanded only about 20\% of DAC’s computational complexity, underscoring the advantages of MDCT-domain coding over direct waveform coding.
Our CFMDCTCodec used only {13\%} of the model parameters {and less than a quarter of the FLOPs} of BigCodec while achieving comparable performance, demonstrating that CFMDCTCodec effectively balanced quality and complexity.
{Compared with FlowDec, CFMDCTCodec achieved a CPU RTF that was only 2\% of FlowDec’s, while requiring merely 0.5\% of its FLOPs.}
This was mainly because FlowDec was built upon a DAC-style waveform codec backbone and applied post-processing on complex-valued STFT representations, whereas CFMDCTCodec operated entirely in the MDCT domain throughout the pipeline, further highlighting the efficiency advantage of MDCT-domain modeling.
WavTokenizer had lower FLOPs, which can be attributed to  {the absence of an additional postprocessing module.}

{We further investigated the algorithmic delay of all compared codecs under the low-bitrate condition.
DAC, BigCodec and FlowDec had finite algorithmic delays of 329\,ms, 372\,ms and 4.82 s, respectively.
However, all the other codecs were based on global operations, and their algorithmic delays therefore varied with the length of the input speech.
For example, MDCTCodec used GRN, and WavTokenizer employed global attention.
CFMDCTCodec inherited GRN from the MDCTCodec backbone and further incorporated utterance-level normalization into the enhancer.
Therefore, under the current configuration, the relatively high algorithmic delay of CFMDCTCodec constituted a limitation of the current system, which will be further addressed in our future work.}

\subsection{Component Analysis for MDCT-Spectral Enhancer}
\label{subsec:Component Analysis for MDCT-Spectral Enhancer} 

The noise-prior-aware CFM-based MDCT-spectral enhancer constitutes a central component of CFMDCTCodec. 
In this subsection, we systematically validated its role in the overall framework and disentangled the contributions of several key components within the enhancer. 
All the following experiments were conducted on the 16-kHz LibriTTS dataset at the low bitrate of 0.65 kbps.

\subsubsection{\textbf{Role Validation via Spectral Visualization}}
\begin{figure}[t]
  \centering
  \includegraphics[width=\linewidth]{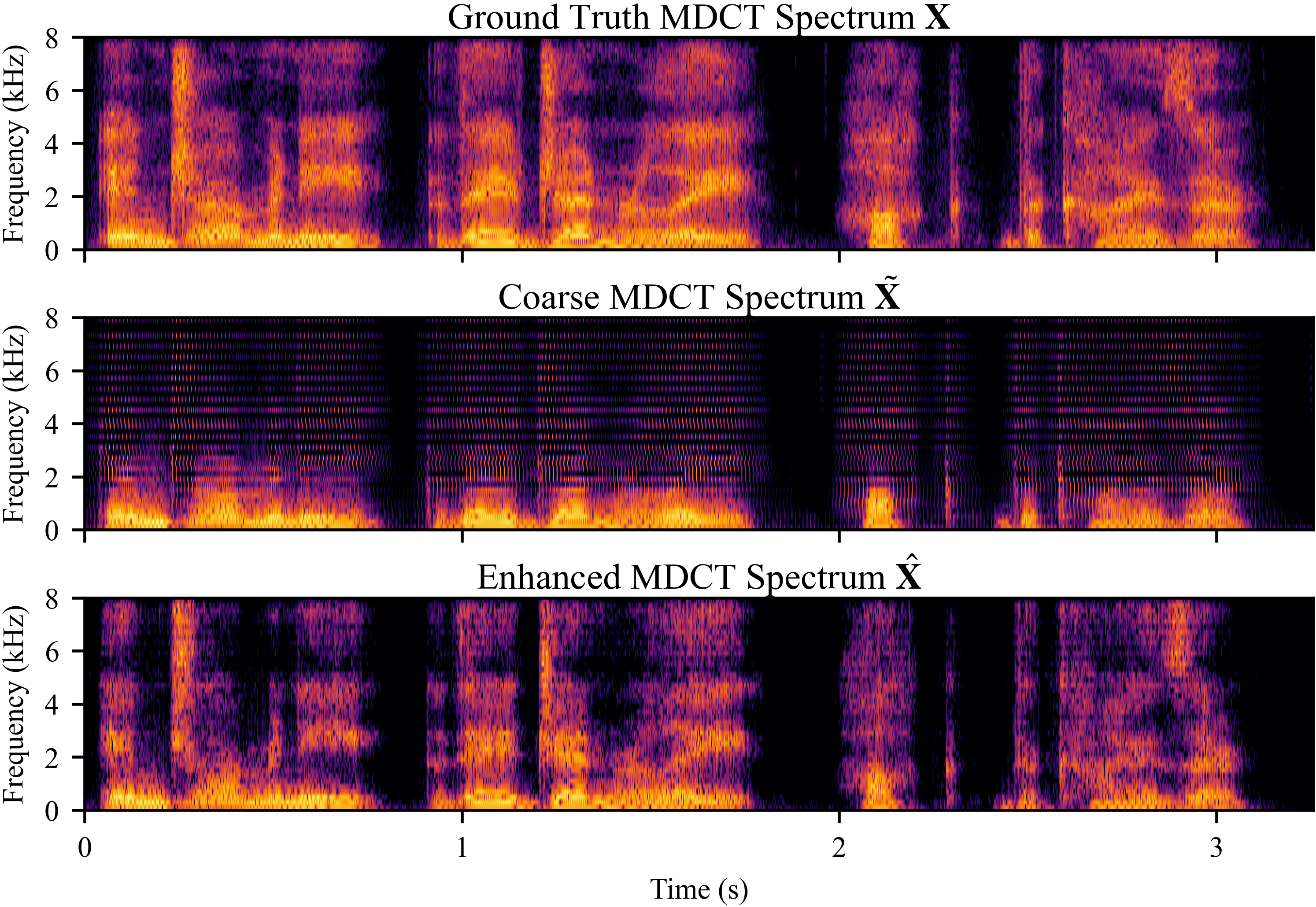}
\caption{Visualization of the ground-truth MDCT spectrum $\mathbf{X}$, and the coarse MDCT spectrum $\tilde{\mathbf{X}}$ and the enhanced MDCT spectrum $\hat{\mathbf{X}}$ generated by CFMDCTCodec at 0.65 kbps for a test utterance in 16-kHz LibriTTS dataset.}
  \label{fig:bwe3_mdct}
\end{figure}

The noise-prior-aware CFM-based MDCT-spectral enhancer was designed to further improve the quality of the coarse MDCT spectrum $\tilde{\mathbf{X}}$ decoded by the single-codebook MDCT-spectral codec, producing an enhanced MDCT spectrum $\hat{\mathbf{X}}$. 
To qualitatively validate the role of this enhancer, we visualized $\tilde{\mathbf{X}}$, $\hat{\mathbf{X}}$, and the ground-truth MDCT spectrum $\mathbf{X}$ for a test utterance. 
As shown in Fig.~\ref{fig:bwe3_mdct}, the coarse MDCT spectrum reconstructed from the ultra-low-bitrate bitstream is severely degraded, with significant loss of fine structures, especially in the high-frequency region where harmonic details are mostly absent. 
{In addition, the coarse spectrum exhibits noticeable horizontal stripe-like patterns, which may be attributed to the severely limited information capacity at extremely low bitrates: the model prioritizes high-energy low-frequency components that are more critical to speech reconstruction, while the high-frequency harmonic structure is represented only through a crude copy-like approximation rather than faithful reconstruction.}
After applying the enhancer, the enhanced MDCT spectrum recovered noticeably clearer harmonic patterns and more coherent spectral trajectories, bringing it substantially closer to the ground truth. 
This qualitative improvement confirms that the proposed enhancer effectively restored MDCT-spectral details and plays a critical role in improving the decoded spectrum quality at low bitrates.

\begin{table}[t]
\large
\centering
\caption{Objective experimental results of CFMDCTCodec and its two ablated variants at 0.65 kbps on the test set of the 16-kHz dataset.}
\label{tab:ablation}
\resizebox{\linewidth}{!}{
\begin{tabular}{l c c c c c c}
\hline

\hline
\textbf{Codec} & \textbf{STOI $\uparrow$} & \textbf{SI-SDR $\uparrow$} & \textbf{SIM $\uparrow$} & {\textbf{LSD $\downarrow$}}& \textbf{DNSMOS $\uparrow$} & \textbf{UTMOS $\uparrow$} \\
\hline
CFMDCTCodec & 0.866 & -3.206 & 0.942 &{1.166} & 3.186 & 3.761 \\
\hline
w/o Range Norm. & 0.860&-3.130& 0.939& {1.191} & 3.003 & 3.437 \\
w/o Adaptive Prior & 0.844&-5.490& 0.935& {1.702} &3.054 & 3.140 \\
\hline

\hline
\end{tabular}}
\end{table}

\begin{figure*}[t]
  \centering
  \includegraphics[width=\textwidth]{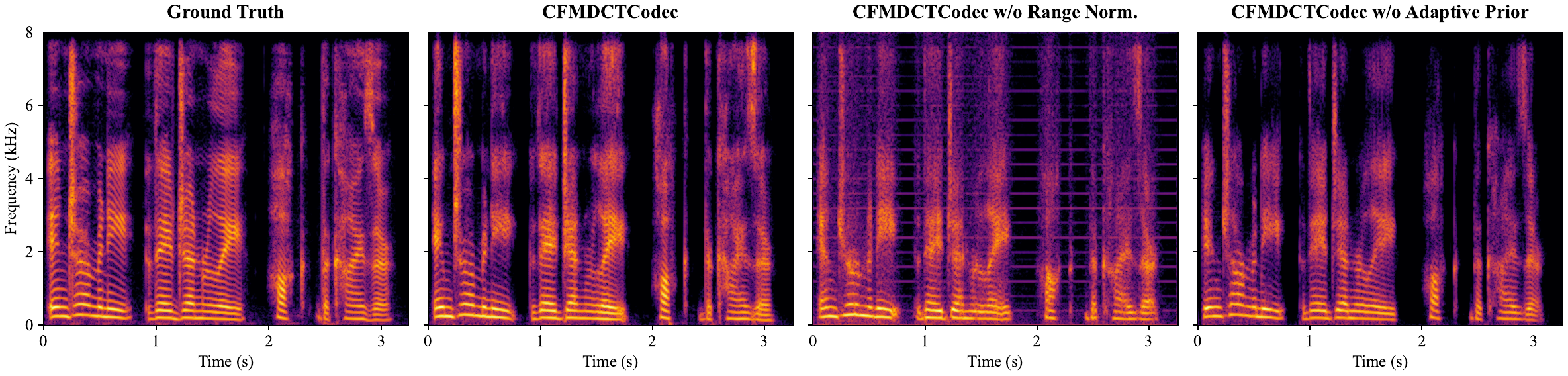} 
  \caption{Spectrograms of the ground-truth speech and the speech generated by CFMDCTCodec and its two ablated variants at 0.65 kbps for a test utterance in 16-kHz LibriTTS dataset.}
  \label{fig:ablation_spec}
\end{figure*}

\subsubsection{\textbf{Effectiveness Analysis of MDCT Range Normalization}}

We applied MDCT range normalization in the enhancer by first mapping the coarse MDCT spectrum to a normalized range before feeding it into the CFM mechanism, and then denormalizing the enhanced output back to the original scale. 
To validate the necessity of this design, we conducted an ablation study by removing the normalization/denormalization pair and training the velocity-field filter directly on raw MDCT spectrum (w/o Range Norm.). 
We evaluated the resulting ablated variant using all quality-related objective metrics, with the results summarized in Table~\ref{tab:ablation}. 
We can see that removing range normalization caused a notable decrease in DNSMOS and UTMOS, reflecting a clear degradation in perceptual quality, whereas the effect on intelligibility remained minor, as suggested by the STOI results. 

\begin{figure}[t]
  \centering
  \includegraphics[width=0.9\linewidth]{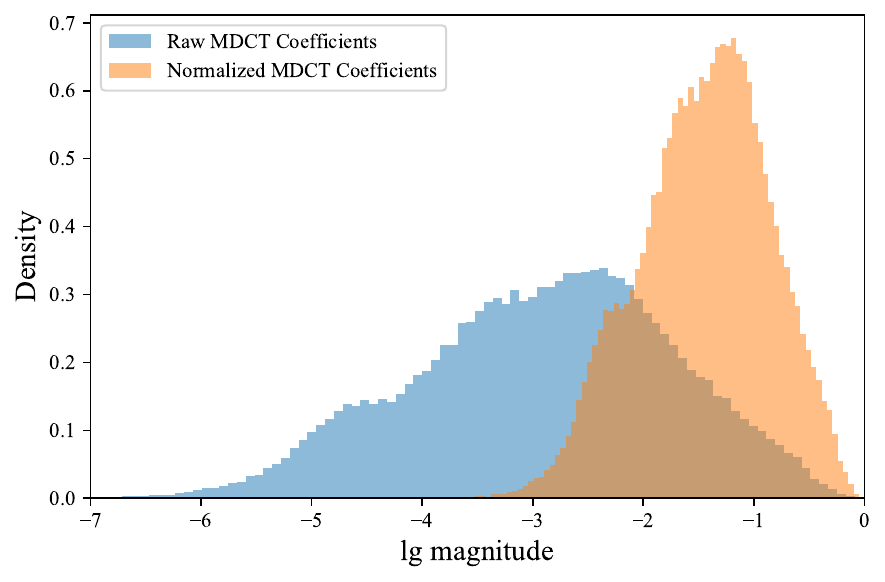}
  \caption{Distribution of MDCT coefficients before (blue) and after (orange) MDCT range normalization for a test utterance in 16-kHz LibriTTS dataset.}
  \label{fig:mdct_hist}
\end{figure}

We further substantiated this finding via visualization. 
Fig.~\ref{fig:mdct_hist} plots the histogram of MDCT coefficients before and after normalization, showing that raw MDCT magnitudes followed a strongly heavy-tailed distribution spanning multiple orders of magnitude, whereas normalization compressed them into a much narrower and better-balanced range. 
This improved numerical conditioning helps stabilize neural optimization in the real-valued MDCT domain.
Consistently, Fig.~\ref{fig:ablation_spec} compares speech spectrograms with and without MDCT range normalization, where the ablated model exhibited pronounced spectral artifacts (e.g., horizontal streaks) that were largely suppressed when normalization was used. 
These results demonstrate that MDCT range normalization is a crucial stabilization mechanism for reliable CFM-based MDCT-spectral enhancement. 

\subsubsection{\textbf{Effectiveness Analysis of Magnitude-Adaptive Noise Prior}}
The magnitude-adaptive noise prior is another core design in the proposed enhancer. 
It adjusts the noise scale according to the magnitude of the coarse MDCT spectrum and is used to construct the CFM initial state. 
To validate its effectiveness, we conducted an ablation study by replacing it with a fixed global noise scale, i.e., fixing $\boldsymbol{\sigma}$ to constant 1.0. 
As shown in Table~\ref{tab:ablation}, this modification led to a consistent degradation in both perceptual and fidelity-related objective metrics, indicating that a non-adaptive prior made the enhancement process less reliable. 
{This effect is further reflected by the LSD results, indicating that the magnitude-adaptive prior plays an important role in restoring fine spectral structures and suppressing frequency-domain distortion.}
Fig.~\ref{fig:ablation_spec} provided clear qualitative evidence. 
With the magnitude-adaptive prior, the enhanced spectrogram became brighter and more energetic, approaching the ground-truth reference, whereas using a fixed noise level tended to yield over-smoothed spectra with attenuated high-frequency details. 
This could be attributed to the fact that the adaptive scaling shaped the CFM initial state to reflect the energy distribution of the coarse MDCT spectrum, which eased the CFM process and ultimately improved MDCT-spectral quality.
Therefore, the magnitude-adaptive noise scaling is crucial for effective CFM-based MDCT-spectral enhancement. 

\begin{table}[t]
\centering
\caption{Objective experimental results of CFMDCTCodec at 0.65 kbps with varying MDCT hop sizes in the MDCT-spectral enhancer on the test set of the 16-kHz dataset.}
\label{tab:fm_hop}
\large
\resizebox{\linewidth}{!}{
\begin{tabular}{c c c c c c c c}
\hline

\hline
\textbf{Hop Size} & \multirow{2}{*}{\textbf{STOI $\uparrow$}}& \multirow{2}{*}{\textbf{{SI-SDR $\uparrow$}}} & \multirow{2}{*}{\textbf{SIM $\uparrow$}} & \multirow{2}{*}{\textbf{{LSD $\downarrow$}}} & \multirow{2}{*}{\textbf{{DNSMOS} $\uparrow$}} & \multirow{2}{*}{\textbf{{UTMOS} $\uparrow$}} & \multirow{2}{*}{\textbf{{FLOPs} $\downarrow$ }}\\
\textbf{(samples)} & & & & & &  \\
\midrule
20  & 0.784 &-14.436&0.918 &{1.097}& 2.306 & 2.174 & 21.49G \\
40 (Aligned) & 0.866 &-3.206&0.942&{1.166}& 3.186 & 3.761 & 11.93G \\
80  & 0.848 &-4.712&0.938&{1.234}& 3.103 & 3.391 & 7.28G \\
160 & 0.826 &-4.811&0.897&{1.308}& 2.718 & 2.576 & 4.90G \\
\hline

\hline
\end{tabular}}
\end{table}

\subsubsection{\textbf{Discussion on MDCT Hop Size}}
In our implementation of CFMDCTCodec, the MDCT used throughout the pipeline adopted a hop size of 40 samples. 
The front-end MDCT-spectral codec operated at this hop size and, under a fixed bitrate configuration, its hop size could not be adjusted. 
In contrast, the back-end MDCT-spectral enhancer was not subject to this constraint. 
We therefore varied the enhancer-side MDCT hop size to 20, 80, and 160 to study its effect. 
Since these settings were no longer aligned with the codec’s original MDCT partitioning, we first converted the coarse MDCT spectrum produced by the codec back to waveform via IMDCT, then re-extracted an MDCT spectrum using the target hop size, and finally fed it into the enhancer for enhancement. 

Table~\ref{tab:fm_hop} summarizes quality- and complexity-related objective metrics of CFMDCTCodec under different MDCT hop sizes in the enhancer. 
Overall, the aligned setting (i.e., hop size $=40$) provided the best balance, achieving the {best overall trade-off between reconstruction quality and computational cost.} 
Increasing the hop size reduced FLOPs substantially but consistently degraded quality, indicating that overly sparse time updates limited the enhancer’s ability to correct fine-grained artifacts. 
Conversely, reducing the hop size to 20 increased computation markedly yet did not translate into better quality. 
These results highlight the importance of time–frequency alignment between the MDCT-spectral codec and enhancer.

\subsubsection{\textbf{{Discussion on Temperature $\tau$}}}
\label{dis_temp}
\begin{figure}[t]
  \centering
  \includegraphics[width=0.9\linewidth]{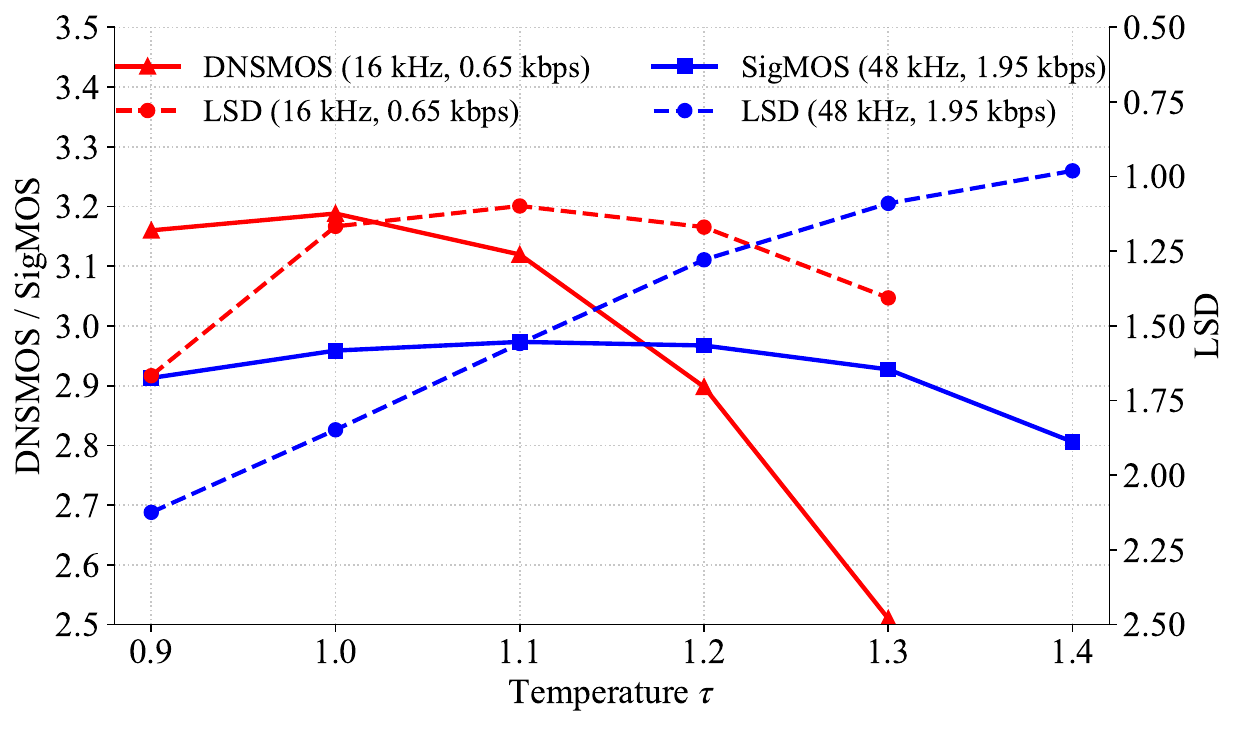}
  \caption{{Impact of the temperature $\tau$ on the performance of CFMDCTCodec at two settings.}}
  \label{fig:temperature_effect}
\end{figure}

{The temperature parameter $\tau$ controls the initial noise scale, balancing sampling stochasticity against fine-detail generation. 
We conducted a temperature sweep at two representative settings (i.e., 16\,kHz / 0.65\,kbps and 48\,kHz / 1.95\,kbps) by varying $\tau$ at inference time with all other settings fixed, and evaluated the results using perceptual quality metrics (i.e., DNSMOS for 16 kHz and SIGMOS for 48 kHz) and spectral distortion (i.e., LSD). 
As shown in Fig.~\ref{fig:temperature_effect}, the sensitivity to the temperature parameter differed across the two sampling rates.
At 16 kHz, DNSMOS peaked at $\tau=1.0$ while LSD remained low; larger $\tau$ noticeably degraded DNSMOS with only marginal improvement in LSD, so we adopted $\tau=1.0$ for this configuration. 
In contrast, for the 48-kHz setting, small temperature values (e.g., $\tau=0.9$ or $1.0$) led to substantially larger LSD, indicating severe frequency-domain distortion.
As suggested by the visualization in Fig.~\ref{fig:bwe3_mdct}, and considering that the 48 kHz setting covers a much wider frequency range, the coarse spectrum in this case may contain more severely weakened high-frequency regions. 
Since the magnitude-adaptive noise prior is derived from the coarse spectral energy, these regions may receive only very weak initial perturbations when $\tau$ is small, thereby limiting the ability of the CFM-based enhancer to regenerate the missing high-frequency components. 
As $\tau$ increased, the global noise scale was enlarged, which helped improve high-frequency restoration. 
This trend is reflected by the continuous reduction in LSD as $\tau$ increased. 
Meanwhile, SIGMOS remained relatively stable over a broad range and only showed a clear degradation at $\tau=1.4$. 
Considering both perceptual stability and spectral fidelity, we chose $\tau=1.3$ for the 48-kHz configuration.
}

\subsection{Training Scheme Discussion}
\label{subsec: Training Scheme Discussion}

When training CFMDCTCodec, we adopted an end-to-end joint training scheme that jointly optimized the MDCT-spectral codec $\phi$ and the enhancer $\theta$. 
As discussed in Section~\ref{subsec:baseline_cmp}, many baselines followed training schemes different from ours. 
For example, FlowDec employed a decoupled two-stage pipeline without adversarial supervision, whereas several other codecs relied on adversarial-based training objectives. 
To validate the effectiveness of the proposed joint optimization scheme in CFMDCTCodec, we constructed two training-strategy variants based on CFMDCTCodec for comparison, as detailed below.

\begin{table}[t]
\large
\centering
\caption{Objective experimental results of CFMDCTCodec at 0.65 kbps for different training schemes on the test set of the 16-kHz dataset.}
\label{tab:training}
\resizebox{\linewidth}{!}{
\begin{tabular}{l c c c c c c}
\hline

\hline
\textbf{Training} & \textbf{STOI $\uparrow$} & \textbf{SI-SDR $\uparrow$} & \textbf{SIM $\uparrow$} & {\textbf{LSD $\downarrow$}} & \textbf{DNSMOS $\uparrow$} & \textbf{UTMOS $\uparrow$} \\
\hline
Joint & 0.866 & -3.206 & 0.942 & {1.166} & 3.186 & 3.761 \\
\hline
Two-Stage & 0.780&-25.928& 0.913 & {1.540} & 2.736 & 2.682 \\
Two-Stage* & 0.788&-25.330& 0.918 & {1.378} & 2.815 & 2.716 \\
\hline

\hline
\end{tabular}}
\end{table}

\begin{itemize}[leftmargin=*, labelsep=0.5em]
  \item \textbf{Two-Stage}: The CFMDCTCodec trained in a decoupled two-stage scheme. 
  It first optimized the MDCT-spectral codec using the spectral reconstruction loss $\mathcal{L}_{\mathrm{spec}}(\phi)$ and the quantization loss $\mathcal{L}_{\mathrm{VQ}}(\phi)$. 
  The codec parameters were then frozen, and the enhancer was trained on the codec outputs using the CFM loss $\mathcal{L}_{\mathrm{CFM}}(\theta)$, i.e., no longer the joint objective $\mathcal{L}_{\mathrm{CFM}}(\phi,\theta)$.
  
  \item \textbf{Two-Stage*}: The CFMDCTCodec trained in a decoupled two-stage scheme augmented with adversarial supervision. 
  On the basis of \textbf{Two-Stage}, it additionally introduced an MDCT-spectral discriminator and an adversarial loss borrowed from \cite{jiang2024mdctcodec} when training the MDCT-spectral codec, aiming to further improve the quality of the decoded coarse MDCT spectrum.
\end{itemize}

The experiments were conducted on the 16-kHz LibriTTS dataset at the low bitrate of 0.65~kbps, and the objective experimental results are summarized in Table~\ref{tab:training}. 
We can see that the proposed end-to-end joint optimization consistently yielded the best performance profile, delivering stronger intelligibility and noticeably higher perceptual quality than the decoupled two-stage variants. 
In contrast, both two-stage pipelines suffered from a severe distortion collapse. 
Introducing an adversarial objective when training the MDCT-spectral codec provided only limited relief and still lagged far behind joint training. 
This may be attributed to the fact that end-to-end joint optimization allowed the MDCT-spectral codec and the enhancer to co-adapt during training, explicitly matching the codec’s coarse output distribution to the enhancer’s input requirements. 
The misalignment between the two-stage components impeded the enhancement process, leading to a noticeable degradation in speech quality, thus further confirming the effectiveness of the end-to-end joint training approach we implemented.
\begin{table}[t]
\large
\centering
\caption{{Objective comparison between MDCT and STFT representations within the CFMDCTCodec framework at 0.65 kbps on the 16-kHz test set.}}
\label{tab:ablation_stft}
\resizebox{\linewidth}{!}{
{
\begin{tabular}{l c c c c c c c c}
\toprule
 {\textbf{Rep.}} & \textbf{STOI $\uparrow$} & \textbf{SI-SDR $\uparrow$} & \textbf{SIM $\uparrow$} & \textbf{LSD $\downarrow$}& \textbf{DNSMOS $\uparrow$} & \textbf{UTMOS $\uparrow$} &  {\textbf{FLOPs $\downarrow$}} &  {\textbf{Param. $\downarrow$}} \\
\midrule
MDCT & 0.866 & -3.206 & 0.942 & 1.166 & 3.186 & 3.761 &  {11.93G} &  {14.61M} \\
STFT &  {0.821} &  {0.233} &  {0.909} &  {1.959} &  {3.101} &  {3.603} &  {1074.00G} &  {41.99M} \\
\bottomrule
\end{tabular}}}
\end{table}

\subsection{{Comparison of Time--Frequency Representations}}
\label{subsec:comp_mdct_stft}
{To validate the necessity of adopting the MDCT representation in CFMDCTCodec's framework, we constructed an equivalent variant based on complex STFT representations for an ablation comparison under the low-bitrate setting of 0.65 kbps at 16 kHz. 
 {The STFT variant differed from CFMDCTCodec only in the time--frequency representation used as the modeling target, i.e., it modeled the STFT spectrum instead of the MDCT spectrum.} 
 {For a fairer complex-valued STFT comparison, we kept the real and imaginary parts as two separate channels, so that the two components at each time--frequency bin could be processed jointly by the codec and enhancer.} 
 {The reconstructed two-channel STFT representation was then converted back into waveform speech via inverse STFT.} 
}
 {The experimental results are shown in Table~\ref{tab:ablation_stft}. 
We can see that, compared with the STFT spectrum, the use of the MDCT spectrum yielded clear advantages on most metrics, indicating higher reconstructed speech quality overall.
In addition, the complex STFT representation required modeling both the real and imaginary components with the 2D codec and CFM-based enhancer, which substantially increased computational complexity and reduced efficiency compared with the real-valued MDCT representation.}
The above experimental results further confirmed the effectiveness of adopting the MDCT spectrum as the modeling target in CFMDCTCodec.

\section{Conclusion} \label{sec:conclusion}

In this paper, we introduced CFMDCTCodec, a neural speech codec designed for high-quality speech coding at low bitrates. 
By operating entirely in the MDCT domain, CFMDCTCodec combines a single-codebook MDCT-spectral codec with a noise-prior-aware CFM-based MDCT-spectral enhancer. 
The front-end codec deeply compresses the MDCT spectrum and provides coarse decoding, while the back-end enhancer boosts the decoding capability by enhancing the coarse MDCT spectrum. 
Within the MDCT-spectral enhancer, we further employed MDCT range normalization and a magnitude-adaptive noise prior to stabilize the CFM-based refinement process. 
When optimizing CFMDCTCodec, we adopted a non-adversarial training scheme that enabled the MDCT-spectral codec and enhancer to co-adapt through the joint optimization of spectral reconstruction, quantization, and CFM objectives. 
Experimental results showed that, at a bitrate of only 0.65~kbps, {it} outperformed competitive baselines and achieved perceptual quality comparable to substantially larger codecs, while using far fewer parameters and lower computational cost. 
In future work, we will investigate pushing CFMDCTCodec to even lower bitrates while further reducing its computational and model complexity, and we will explore low algorithmic delay, streaming-friendly configurations to better support real-time deployment.

\bibliographystyle{IEEEtran}
\bibliography{mybib}
\end{document}